%% file: main.tex
\pdfoutput=1
\documentclass[12pt,a4paper]{article}
\usepackage{ifthen} % for conditional statements
\newboolean{pdflatex}
\setboolean{pdflatex}{true} % False for eps figures 

\newboolean{articletitles}
\setboolean{articletitles}{true} % False removes titles in references

\newboolean{uprightparticles}
\setboolean{uprightparticles}{false} %True for upright particle symbols

\def\paperauthors{LHCb collaboration} % Leave as is for PAPER, CONF and FIGURE
\def\paperasciititle{Inclusive B-meson flavour-tagging algorithm at LHCb} % Set ASCII title here !! MAKE sure it's only ASCII characters !! 
\def\papertitle{Inclusive $B$-meson flavour-tagging algorithm at LHCb} % Latex formatted title
\def\paperkeywords{{High Energy Physics}, {LHCb}} % Comma separated list
\def\papercopyright{\the\year\ CERN for the benefit of the LHCb collaboration} % new since 9/Apr/2018
\def\paperlicence{CC BY 4.0 licence}
\def\paperlicenceurl{https://creativecommons.org/licenses/by/4.0/}

\newif\ifEnableSectionTOCLinks
\EnableSectionTOCLinksfalse % deactivated

\input{preamble}
\usepackage{longtable} % only for template; not usually to be used in PAPERs

\begin{document}

%%%%%%%%%%%%%%%%%%%%%%%%%
%%%%% Title     %%%%%%%%%
%%%%%%%%%%%%%%%%%%%%%%%%%
\renewcommand{\thefootnote}{\fnsymbol{footnote}}
\setcounter{footnote}{1}

% %%%%%%% CHOOSE TITLE PAGE--------
%\onecolumn
%\input{title-LHCb-INT}
%\input{title-LHCb-ANA}
%\input{title-LHCb-CONF}
%\input{title-LHCb-FIGURE}
\input{title-LHCb-PAPER}

%\twocolumn
% %%%%%%%%%%%%% ---------

\renewcommand{\thefootnote}{\arabic{footnote}}
\setcounter{footnote}{0}

%%%%%%%%%%%%%%%%%%%%%%%%%%%%%%%%
%%%%%  Table of Content   %%%%%%
%%%%%%%%%%%%%%%%%%%%%%%%%%%%%%%%
%%%% Uncomment if desired
%\tableofcontents

\cleardoublepage

%%%%%%%%%%%%%%%%%%%%%%%%%
%%%%% Main text %%%%%%%%%
%%%%%%%%%%%%%%%%%%%%%%%%%

\pagestyle{plain} % restore page numbers for the main text
\setcounter{page}{1}
\pagenumbering{arabic}

%% Uncomment during review phase. 
%% Comment before a final submission.
%\linenumbers

%% This is the main body
%% It is useful to have a single file so comments are not missed in overleaf.
\input{body}

% Do not include this in any draft (just for information in the template)
% \input{acknowledgements_intro}
% Comment this in for paper drafts; do not include this in analysis note, conference and figure reports
\input{acknowledgements}

%\input{supplementary}

%\input{appendix}

% This should be taken out in the final paper
%\input{supplementary-app}

\addcontentsline{toc}{section}{References}
%\setboolean{inbibliography}{true}
\bibliographystyle{LHCb}
\bibliography{main,standard,LHCb-PAPER,LHCb-CONF,LHCb-DP,LHCb-TDR}

\newpage
\input{Authorship_LHCb-PAPER-2025-024}

\end{document}

%% file: preamble.tex
\usepackage[top=1in, bottom=1.25in, left=1in, right=1in]{geometry}

\usepackage{microtype}
\usepackage{lineno}  % for line numbering during review
\usepackage{xspace} % To avoid problems with missing or double spaces after
\usepackage{caption} %these three command get the figure and table captions automatically small
\usepackage{booktabs}
\usepackage{multirow}

\usepackage{graphicx}  % to include figures (can also use other packages)
\usepackage{color}
\usepackage{colortbl}
\graphicspath{{./figs/}} % Make Latex search fig subdir for figures
\usepackage{amsmath} % Adds a large collection of math symbols
\usepackage{amssymb}
\usepackage{amsfonts}
\usepackage{upgreek} % Adds in support for greek letters in roman typeset

\newcommand*\patchAmsMathEnvironmentForLineno[1]{%
\expandafter\let\csname old#1\expandafter\endcsname\csname #1\endcsname
\expandafter\let\csname oldend#1\expandafter\endcsname\csname
end#1\endcsname
 \renewenvironment{#1}%
   {\linenomath\csname old#1\endcsname}%
   {\csname oldend#1\endcsname\endlinenomath}%
}
\newcommand*\patchBothAmsMathEnvironmentsForLineno[1]{%
  \patchAmsMathEnvironmentForLineno{#1}%
  \patchAmsMathEnvironmentForLineno{#1*}%
}
\AtBeginDocument{%
\patchBothAmsMathEnvironmentsForLineno{equation}%
\patchBothAmsMathEnvironmentsForLineno{align}%
\patchBothAmsMathEnvironmentsForLineno{flalign}%
\patchBothAmsMathEnvironmentsForLineno{alignat}%
\patchBothAmsMathEnvironmentsForLineno{gather}%
\patchBothAmsMathEnvironmentsForLineno{multline}%
\patchBothAmsMathEnvironmentsForLineno{eqnarray}%
}

\usepackage{hyperxmp}

\usepackage[pdftex,
            pdfauthor={\paperauthors},
            pdftitle={\paperasciititle},
            pdfkeywords={\paperkeywords},
            pdfcopyright={Copyright (C) \papercopyright},
            pdflicenseurl={\paperlicenceurl}]{hyperref}

\usepackage[colorinlistoftodos,textsize=scriptsize]{todonotes}

\usepackage[bottom,flushmargin,hang,multiple]{footmisc}

\usepackage[all]{hypcap} % Internal hyperlinks to floats.

\input{lhcb-symbols-def} % Add in the predefined LHCb symbols

\hypersetup{
  colorlinks   = true, %Colours links instead of ugly boxes
  urlcolor     = blue, %Colour for external hyperlinks
  linkcolor    = blue, %Colour of internal links
  citecolor    = red   %Colour of citations
}

\ifEnableSectionTOCLinks
    \usepackage[explicit]{titlesec} % to change headings
    
    \let\oldcontentsline\contentsline
    \renewcommand

    \titleformat{\section}{\normalfont\Large\bf}{\hyperlink{tocsection.\thesection}{{\thesection} \parbox[t]{\dimexpr\textwidth-1pc}{#1}}}{1pc}{}

    \titleformat{\subsection}{\normalfont\bf}{\hyperlink{tocsubsection.\thesubsection}{{\thesubsection} \parbox[t]{\dimexpr\textwidth-1pc}{#1}}}{1pc}{}

    \titleformat{name=\section,numberless}[display]{}{}{0pt}{\normalfont\Huge\bfseries #1}
\fi

\usepackage{cite} % Allows for ranges in citations
\usepackage{mciteplus}

%% file: lhcb-symbols-def.tex
\usepackage{xspace} 
\usepackage{upgreek}

\def\lhcb   {\mbox{LHCb}\xspace}

\def\belle  {\mbox{Belle}\xspace}

\def\MagUp {\mbox{\em Mag\kern -0.05em Up}\xspace}

\ifthenelse{\boolean{uprightparticles}}%
{

 \def\Pmu         {\ensuremath{\upmu}\xspace}

 \def\Ppi         {\ensuremath{\uppi}\xspace}

 \def\Ppsi        {\ensuremath{\uppsi}\xspace}

 \def\PDelta      {\ensuremath{\Delta}\xspace}                 
 \def\PXi         {\ensuremath{\Xi}\xspace}                 
 \def\PLambda     {\ensuremath{\Lambda}\xspace}                 
 \def\PSigma      {\ensuremath{\Sigma}\xspace}                 
 \def\POmega      {\ensuremath{\Omega}\xspace}                 
 \def\PUpsilon    {\ensuremath{\Upsilon}\xspace}
 \let\oldPi\Pi
 \def\PPi         {\ensuremath{\oldPi}\xspace}

 \def\PB      {\ensuremath{\mathrm{B}}\xspace}                 
 \def\PD      {\ensuremath{\mathrm{D}}\xspace}                 
 \def\PJ      {\ensuremath{\mathrm{J}}\xspace}                 
 \def\PK      {\ensuremath{\mathrm{K}}\xspace}                 
 \def\Pb      {\ensuremath{\mathrm{b}}\xspace}                 
 \def\Pc      {\ensuremath{\mathrm{c}}\xspace}                 
 \def\Pd      {\ensuremath{\mathrm{d}}\xspace}                 

 \def\Pp      {\ensuremath{\mathrm{p}}\xspace}                 

 \def\Ps      {\ensuremath{\mathrm{s}}\xspace}

 \def\thebaroffset{0.0em}
}
{

 \def\Pmu         {\ensuremath{\mu}\xspace}

 \def\Ppi         {\ensuremath{\pi}\xspace}

 \def\Ppsi        {\ensuremath{\psi}\xspace}                 
                  
 \mathchardef\PDelta="7101
 \mathchardef\PXi="7104
 \mathchardef\PLambda="7103
 \mathchardef\PSigma="7106
 \mathchardef\POmega="710A
 \mathchardef\PUpsilon="7107
 \mathchardef\PPi="7105
 \def\PB      {\ensuremath{B}\xspace}                 
 \def\PD      {\ensuremath{D}\xspace}                 
 \def\PJ      {\ensuremath{J}\xspace}                 
 \def\PK      {\ensuremath{K}\xspace}                 
 \def\Pb      {\ensuremath{b}\xspace}                 
 \def\Pc      {\ensuremath{c}\xspace}                 
 \def\Pd      {\ensuremath{d}\xspace}                 

 \def\Pp      {\ensuremath{p}\xspace}                 

 \def\Ps      {\ensuremath{s}\xspace}

 \def\thebaroffset{0.18em}
}
\newcommand{\offsetoverline}[2][\thebaroffset]{\kern #1\overline{\kern -#1 #2}}%

\makeatletter
\ifcase \@ptsize \relax% 10pt
  \newcommand{\miniscule}{\@setfontsize\miniscule{4}{5}}% \tiny: 5/6
\or% 11pt
  \newcommand{\miniscule}{\@setfontsize\miniscule{5}{6}}% \tiny: 6/7
\or% 12pt
  \newcommand{\miniscule}{\@setfontsize\miniscule{5}{6}}% \tiny: 6/7
\fi
\makeatother

\DeclareRobustCommand{\optbar}[1]{\shortstack{{\miniscule (\rule[.5ex]{1.25em}{.18mm})}
  \\ [-.7ex] $#1$}}

\def\mumu       {{\ensuremath{\Pmu^+\Pmu^-}}\xspace}

\def\squark    {{\ensuremath{\Ps}}\xspace}

\def\cquark    {{\ensuremath{\Pc}}\xspace}

\def\bquark    {{\ensuremath{\Pb}}\xspace}
\def\bquarkbar {{\ensuremath{\overline \bquark}}\xspace}
\def\bbbar     {{\ensuremath{\bquark\bquarkbar}}\xspace}

\def\pion   {{\ensuremath{\Ppi}}\xspace}

\def\pip    {{\ensuremath{\pion^+}}\xspace}
\def\pim    {{\ensuremath{\pion^-}}\xspace}

\def\kaon    {{\ensuremath{\PK}}\xspace}
\def\KorKbar {\kern \thebaroffset\optbar{\kern -\thebaroffset \PK}{}\xspace}

\def\Kp      {{\ensuremath{\kaon^+}}\xspace}
\def\Km      {{\ensuremath{\kaon^-}}\xspace}

\def\KS      {{\ensuremath{\kaon^0_{\mathrm{S}}}}\xspace}

\def\Kstarz  {{\ensuremath{\kaon^{*0}}}\xspace}

\def\Kstar   {{\ensuremath{\kaon^*}}\xspace}

\def\D       {{\ensuremath{\PD}}\xspace}

\def\DorDbar {\kern \thebaroffset\optbar{\kern -\thebaroffset \PD}\xspace}
\def\Dz      {{\ensuremath{\D^0}}\xspace}

\def\Dp      {{\ensuremath{\D^+}}\xspace}
\def\Dm      {{\ensuremath{\D^-}}\xspace}

\def\DpDm    {\ensuremath{\Dp {\kern -0.16em \Dm}}\xspace}

\def\Dsm     {{\ensuremath{\D^-_\squark}}\xspace}

\def\B       {{\ensuremath{\PB}}\xspace}
\def\Bbar    {{\ensuremath{\offsetoverline{\PB}}}\xspace}

\def\BorBbar {\kern \thebaroffset\optbar{\kern -\thebaroffset \PB}\xspace}
\def\Bz      {{\ensuremath{\B^0}}\xspace}

\def\Bd      {{\ensuremath{\B^0}}\xspace}

\def\BdorBdbar {\kern \thebaroffset\optbar{\kern -\thebaroffset \Bd}\xspace}

\def\Bs      {{\ensuremath{\B^0_\squark}}\xspace}

\def\BsorBsbar {\kern \thebaroffset\optbar{\kern -\thebaroffset \Bs}\xspace}

\def\jpsi     {{\ensuremath{{\PJ\mskip -3mu/\mskip -2mu\Ppsi}}}\xspace}

\def\Y#1S{\ensuremath{\PUpsilon{(#1S)}}\xspace}

\def\LorLbar     {\kern \thebaroffset\optbar{\kern -\thebaroffset \PLambda}\xspace}

\newcommand{\decay}[2]{\mbox{\ensuremath{#1\!\to #2}}\xspace} 

\def\to                 {\ensuremath{\rightarrow}\xspace}

\def\CP                {{\ensuremath{C\!P}}\xspace}

\def\AT#1     {\ensuremath{A_{\mathrm{T}}^{#1}}\xspace}           % 2

\def\C#1      {\ensuremath{\mathcal{C}_{#1}}\xspace}                       % 9
\def\Cp#1     {\ensuremath{\mathcal{C}_{#1}^{'}}\xspace}                    % 7
\def\Ceff#1   {\ensuremath{\mathcal{C}_{#1}^{\mathrm{(eff)}}}\xspace}        % 9  
\def\Cpeff#1  {\ensuremath{\mathcal{C}_{#1}^{'\mathrm{(eff)}}}\xspace}       % 7
\def\Ope#1    {\ensuremath{\mathcal{O}_{#1}}\xspace}                       % 2
\def\Opep#1   {\ensuremath{\mathcal{O}_{#1}^{'}}\xspace}                    % 7

\newcommand{\nospaceunit}[1]{\ensuremath{\text{#1}}}       
\newcommand{\aunit}[1]{\ensuremath{\text{\,#1}}}       
\newcommand{\tev}{\aunit{Te\kern -0.1em V}\xspace}
\newcommand{\gev}{\aunit{Ge\kern -0.1em V}\xspace}
\newcommand{\mev}{\aunit{Me\kern -0.1em V}\xspace}
\newcommand{\kev}{\aunit{ke\kern -0.1em V}\xspace}
\newcommand{\ev}{\aunit{e\kern -0.1em V}\xspace}
 
\newcommand{\mevc}{\ensuremath{\aunit{Me\kern -0.1em V\!/}c}\xspace}
\newcommand{\gevc}{\ensuremath{\aunit{Ge\kern -0.1em V\!/}c}\xspace}
\newcommand{\mevcc}{\ensuremath{\aunit{Me\kern -0.1em V\!/}c^2}\xspace}
\newcommand{\gevcc}{\ensuremath{\aunit{Ge\kern -0.1em V\!/}c^2}\xspace}
\def\cm   {\aunit{cm}\xspace}

\def\mum  {\ensuremath{\,\upmu\nospaceunit{m}}\xspace}

\def\sec  {\ensuremath{\aunit{s}}\xspace}

\newcommand{\chisq}{\ensuremath{\chi^2}\xspace}

\def\gsim{{~\raise.15em\hbox{$>$}\kern-.85em
          \lower.35em\hbox{$\sim$}~}\xspace}
\def\lsim{{~\raise.15em\hbox{$<$}\kern-.85em
          \lower.35em\hbox{$\sim$}~}\xspace}

\def\sPlot{\mbox{\em sPlot}\xspace}

\def\pt         {\ensuremath{p_{\mathrm{T}}}\xspace}

\def\ptot       {\ensuremath{p}\xspace}

\def\evtgen     {\mbox{\textsc{EvtGen}}\xspace}

\def\geant      {\mbox{\textsc{Geant4}}\xspace}

\def\photos     {\mbox{\textsc{Photos}}\xspace}

\def\pythia     {\mbox{\textsc{Pythia}}\xspace}

\def\tell1  {TELL1\xspace}
\def\ukl1   {UKL1\xspace}

\newcommand{\eg}{\mbox{\itshape e.g.}\xspace}
\newcommand{\ie}{\mbox{\itshape i.e.}\xspace}

\newcommand{\lhcborcid}[1]{\href{https://orcid.org/#1}{\hspace*{0.1em}\raisebox{-0.45ex}{\includegraphics[width=1em]{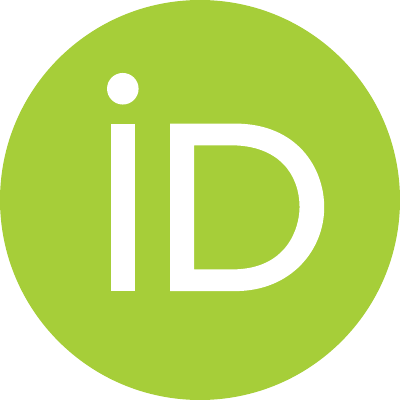}}}}

%% file: title-LHCb-PAPER.tex
% ===============================================================================
% Purpose: LHCb-PAPER journal paper title page template
% Author: 
% Created on: 2010-09-25
% ===============================================================================

%%%%%%%%%%%%%%%%%%%%%%%%%
%%%%%  TITLE PAGE  %%%%%%
%%%%%%%%%%%%%%%%%%%%%%%%%
\begin{titlepage}
\pagenumbering{roman}

% Header ---------------------------------------------------
\vspace*{-1.5cm}
\centerline{\large EUROPEAN ORGANIZATION FOR NUCLEAR RESEARCH (CERN)}
\vspace*{1.5cm}
\noindent
\begin{tabular*}{\linewidth}{lc@{\extracolsep{\fill}}r@{\extracolsep{0pt}}}
\ifthenelse{\boolean{pdflatex}}% Logo format choice
{\vspace*{-1.5cm}\mbox{\!\!\!\includegraphics[width=.14\textwidth]{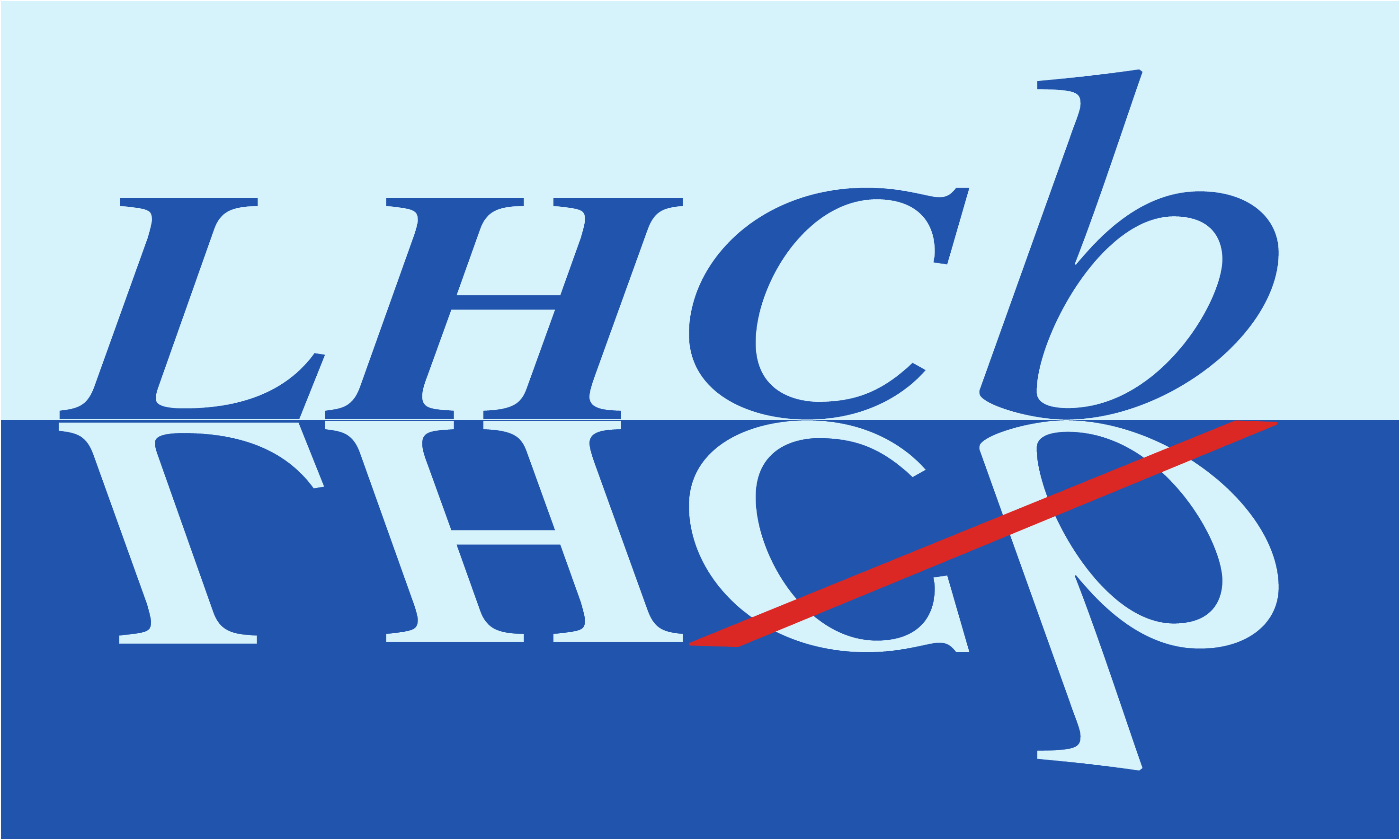}} & &}%
{\vspace*{-1.2cm}\mbox{\!\!\!\includegraphics[width=.12\textwidth]{figs/lhcb-logo.pdf}} & &}%
\\
 & & CERN-EP-2025-168 \\  % ID 
 & & LHCb-PAPER-2025-024 \\  % ID 
 & & November 13, 2025 \\ % Date - Can also hardwire e.g.: 23 March 2010
 & & \\
% not in paper \hline
\end{tabular*}

\vspace*{4.0cm}

% Title --------------------------------------------------
{\normalfont\bfseries\boldmath\huge
\begin{center}
% DO NOT EDIT HERE. Instead edit macro in main.tex to keep metadata correct
  \papertitle 
\end{center}
}

\vspace*{2.0cm}

% Authors -------------------------------------------------
\begin{center}
%In the footnote, replace 'paper' by 'Letter' in case of submission to PRL or PLB 
% Edit macro in main.tex to keep metadata correct
\paperauthors\footnote{Authors are listed at the end of this paper.}
\end{center}

\vspace{\fill}

% Abstract -----------------------------------------------
\begin{abstract}
  \noindent
 A new algorithm is developed to identify the flavour of neutral $B$ mesons at production in $pp$ collisions by utilising all tracks from the hadronisation process. The algorithm is calibrated separately for $B^0$ and $B^{0}_{s}$ mesons using $B^{0}\to J/\psi K^{+}\pi^-$ and 
$B^{0}_{s}\to D_{s}^{-}\pi^+$ decays from  $pp$ collision data collected by the LHCb experiment at a centre-of-mass energy of 13\,TeV. This new algorithm improves the tagging power by 35\% for $B^{0}$ mesons and 20\% for $B^{0}_{s}$ mesons when compared to the combined performance of the existing LHCb flavour-tagging algorithms.
\end{abstract}

\vspace*{2.0cm}

\begin{center}
  Published in JHEP 11 (2025) 041 %/
%   Phys.~Rev.~D /
%   Phys.~Rev.~Lett. /
% %  Phys.~Lett.~B /
%   Eur.~Phys.~J.~C /
%   %  Nucl.~Phys.~B /
%   Chin.~Phys.~C /
%   Nature~Physics /
%   sciPost~Physics /
%   J. Instr. /
%   Instruments 
\end{center}

\vspace{\fill}

{\footnotesize 
% Edit macro in main.tex to keep metadata correct
\centerline{\copyright~\papercopyright. \href{\paperlicenceurl}{\paperlicence}.}}
\vspace*{2mm}

\end{titlepage}

%%%%%%%%%%%%%%%%%%%%%%%%%%%%%%%%
%%%%%  EOD OF TITLE PAGE  %%%%%%
%%%%%%%%%%%%%%%%%%%%%%%%%%%%%%%%

%  empty page follows the title page ----
\newpage
\setcounter{page}{2}
\mbox{~}
%\newpage
%
%% Author List ----------------------------
%%  You need to get a new author list!
%\input{LHCb_authorlist.tex}
%
%The author list for journal publications is provided by the Membership Committee shortly after 'approval to go to paper' has been given.
%%It will be made available on the page
%%\verb!http://www.physik.uzh.ch/~strauman/forMemCo/LHCb-PAPER-XXXX-XXX/! .
%It will be sent to you by email shortly after a paper number has beens assigned.
%The author list should be included already at first circulation, 
%to allow new members of the collaboration to verify whether they have been included correctly.
%Occasionally a misspelled name is corrected or associated institutions become full members.
%In that case, a new author list will be sent to you.
%In case line numbering doesn't work well after including the authorlist, try moving the \verb!\bigskip! after the last author to a separate line.
%
%
%The authorship for Conference Reports should be ``The LHCb
%  collaboration'', with a footnote giving the name(s) of the contact
%  author(s), but without the full list of collaboration names.

%% file: body.tex
\section{Introduction}
\label{sec:introduction}
Measurements of flavour oscillation frequencies~\cite{LHCb-PAPER-2015-031, LHCb-PAPER-2021-005} and time-dependent charge-parity
(\CP) asymmetries of neutral \PB mesons~\cite{LHCb-PAPER-2023-013,LHCb-PAPER-2023-016} require knowledge of the signal \PB-meson flavour at production. The experimental technique to determine this information at
collider experiments is referred to as flavour tagging.  

At the \lhcb experiment, dedicated algorithms are used to infer the production flavour by exploiting the production and decay mechanisms of \PB mesons in $\Pp\Pp$ collisions. 
They are classified into two types of algorithms, opposite-side (OS) and same-side (SS) taggers. The OS taggers~\cite{LHCb-PAPER-2015-027,LHCb-PAPER-2011-027} 
exploit the fact that \Pb quarks are predominantly produced in \bbbar pairs, and thus the flavour at production of the reconstructed \PB meson is opposite to that of the other \Pb hadron in the event, referred to as the opposite-side \Pb hadron. Therefore, flavour tagging can be performed using the decay products of the accompanying \Pb hadron in the event. The five OS taggers in use at \lhcb  are based on muons, electrons, kaons, charm decays and the charge of the \PB decay vertex.  The SS taggers~\cite{LHCb-PAPER-2015-056, LHCb-PAPER-2016-039} use charged particles produced in the hadronisation process in association with the reconstructed signal \PB meson. For \Bs mesons, the SS tagger uses a charged kaon and for \Bz mesons, a charged pion or a proton. The charge of these particles indicates the \Pb-quark content of the \PB meson. Information from SS and OS algorithms is usually combined in flavour-tagged analyses to increase the measurement precision. 
Flavour-tagging algorithms provide an event-by-event tagging decision as well as a mistag estimation $\eta$ that gives the probability of a false decision, which needs to be calibrated on control samples of flavour-specific decays to determine the mistag probability $\omega(\eta)$ of the event.

The performance of a flavour-tagging algorithm is measured by its tagging power $\varepsilon_\text{eff}$, which is defined as
\begin{align}
 \varepsilon_\text{eff} = \varepsilon_\text{tag}\langle D^2\rangle,
    &&  D=1-2\omega,
   \label{eq:taggingmetrics}
\end{align}
where $\varepsilon_\text{tag}$ denotes the fraction of events that have a tagging decision and $D$ is the per-event dilution.
Mistagged candidates reduce the oscillation amplitudes  and  therefore the statistical power of flavour-tagged data.
Notably, the dilution factor $\langle D^2\rangle$ is  obtained by averaging $D^2$ over all events.
The tagging power quantifies the effective statistical sample size, making improved flavour tagging essential to maximise the precision of the oscillation-related observables.
In the case of weighted events $\varepsilon_\text{tag}$ and $\langle D^2 \rangle$ are computed as
\begin{align}
    \varepsilon_\text{tag} &= \frac{\sum_j w_j}{\sum_i w_i}, 
    && \langle D^2 \rangle = \frac{\sum_j w_j(1-2\omega_j)^2}{\sum_j w_j},
\end{align}
where $w$ is the per-event weight and the index $i$ runs over all events, while the index $j$ runs over the subset of tagged events.

This paper describes a new approach called {\it inclusive flavour tagging} (IFT), which exploits the available
information in a $\Pp\Pp$ collision event containing a \bbbar pair at a more global level than the SS and OS taggers~\cite{likhomanenko}. Instead of searching for information that matches particular underlying physics processes as the SS and OS tagging algorithms do, the IFT includes information from the majority of charged-particle tracks produced in an event.  A similar technique, developed for electron-positron collisions, was recently implemented at the \belle~II experiment~\cite{Belle-II:2024lwr} and showed improvement over existing methods.

The IFT presented here is based on a deep neural network architecture DeepSets~\cite{zaheer2018deep}. Two IFT algorithms, one for \Bz mesons and one for \Bs mesons, are trained on simulated \mbox{$\Bz\to\jpsi[\to\mu^{+}\mu^{-}]\Kstar(892)^0\left[\to \Kp\pim\right]$} and \mbox{$\Bs\to\jpsi[\to\mu^{+}\mu^{-}]\phi(1020)\left[\to \Kp\Km\right]$} decays, respectively.\footnote{The inclusion of charge-conjugate processes is implied throughout this paper, unless otherwise noted.}$^{,}$\footnote{
For simplicity, the resonances $\Kstar(892)^0$ and $\phi(1020)$ are referred to as \Kstarz and $\phi$ in the following.} Each algorithm returns an independent event-by-event probability $y\in [0,1]$, from which a tagging decision $d$ and a mistag prediction $\eta$ are set according to the value of $y$ as 
\begin{equation}  
d =  \begin{cases}
       +1 &\quad\text{if}\quad y>0.5 \, ,\\
       \phantom{+}0 &\quad\text{if}\quad y=0.5 \, ,\\
       -1 &\quad\text{if}\quad y<0.5\, ,
     \end{cases}
\end{equation}
where $+1$ corresponds to a \PB meson and $-1$ to \Bbar meson,
and 
\begin{equation}  
\eta = 
     \begin{cases}
       1-y &\quad\text{if}\quad y>0.5\, ,\\
       y &\quad\text{if}\quad y\leq0.5\, .
     \end{cases}
\end{equation}

The mistag probabilities are calibrated on \mbox{$\Bz\to\jpsi\Kp\pim$} decays,  where the kaon-pion pair lies within the mass range of the $\Kstar(892)^0$ resonance, and on \mbox{$\Bs\to\Dsm\pip$} decays using $\Pp\Pp$ collision data collected by the \lhcb experiment at a centre-of-mass energy of 13\tev during Run~2. This calibration is essential because the classifiers are trained on simulated samples, which do not perfectly replicate the characteristics of \lhcb data.  The IFT is validated on \mbox{$\Bz\to\jpsi\KS\left[\to\pip\pim\right]$} decays and \mbox{$\Bs\to\jpsi \Kp\Km$} decays in the vicinity of the $\phi$ resonance.

All studies are performed independently on the samples collected in the years 2016, 2017 and 2018 to account for possible differences between the data collected in varying data-taking conditions.

\section{Detector and simulation}
\label{sec:detector}
The \lhcb detector~\cite{LHCb-DP-2008-001,LHCb-DP-2014-002} is a single-arm forward
spectrometer covering the \mbox{pseudorapidity} range between 2 and 5,
designed for the study of particles containing \bquark or \cquark
quarks. The detector used for these studies includes a high-precision tracking system
consisting of a silicon-strip vertex detector (VELO) surrounding the $pp$
interaction region~\cite{LHCb-DP-2014-001}, a large-area silicon-strip detector (TT) located
upstream of a dipole magnet with a bending power of about
$4{\mathrm{\,T\,m}}$, and three stations of silicon-strip detectors and straw
drift tubes~\cite{LHCb-DP-2017-001}
placed downstream of the magnet.
The tracking system provides a measurement of the momentum, \ptot, of charged particles with
a relative uncertainty that varies from 0.5\% at low momentum to 1.0\% at 200\gevc.
The minimum distance of a track to a primary $pp$ collision vertex (PV), the impact parameter (IP), 
is measured with a resolution of $(15+29/\pt)\mum$,
where \pt is the component of the momentum transverse to the beam, in\,\gevc.
Different types of charged hadrons are distinguished using information
from two ring-imaging Cherenkov detectors~\cite{LHCb-DP-2012-003}. 
Photons, electrons and hadrons are identified by a calorimeter system consisting of
scintillating-pad and preshower detectors, an electromagnetic
and a hadronic calorimeter. Muons are identified by a
system composed of alternating layers of iron and multiwire
proportional chambers~\cite{LHCb-DP-2012-002}.

The online event selection is performed by a trigger~\cite{LHCb-DP-2012-004}, 
which consists of a hardware stage, based on information from the calorimeter and muon
systems, followed by a software stage, which applies a full event
reconstruction. At the hardware trigger stage, events are required to have a muon with high \pt or a  hadron, photon or electron with high transverse energy in the calorimeters. Following the hardware trigger stage, different software trigger algorithms are applied depending on the decay mode, following the strategy of Refs.~\cite{LHCb-PAPER-2021-005,LHCb-PAPER-2023-013,LHCb-PAPER-2023-016}.

In the modes containing a \jpsi meson, the trigger relies on reconstructing the \mbox{$\jpsi\to\mumu$} decay and identifying candidates consistent with a \jpsi meson displaced from the PV associated with the signal \PB-meson decay. 
In the \mbox{$\Bs\to\Dsm\pip$} decay mode, the first-stage software trigger applies a multivariate algorithm to select either a good-quality track with a high \pt or a good-quality two-track secondary vertex. In the second stage, the software trigger selects candidates consistent
with a $b$-hadron decay topology, with a two-, three-, or four-track vertex with a sizeable
\pt and a significant displacement from any PV. 
Note that the hadron trigger applies stricter \pt requirements than the dimuon trigger, leading to a \pt spectrum that is shifted toward higher values.

Simulated events are used to train the flavour-tagging algorithms, to model the signal mass and the decay-time distributions, to verify the analysis procedure and to study systematic effects. In the simulation, $pp$ collisions are generated using \pythia~\cite{Sjostrand:2007gs,*Sjostrand:2006za} with a specific \lhcb configuration~\cite{LHCb-PROC-2010-056}. Decays of unstable particles are described by \evtgen~\cite{Lange:2001uf}, in which final-state radiation is generated using \photos~\cite{davidson2015photos}.  The interaction of the generated particles with the detector, and its response,  are implemented using the \geant toolkit~\cite{Allison:2006ve, *Agostinelli:2002hh} as described in Ref.~\cite{LHCb-PROC-2011-006}. 

\section{Inclusive flavour-tagger training}
\label{sec:training}
In $\Pp\Pp$ collision events, the tracks observed after excluding those associated with the signal \PB-meson decay can be grouped into four categories. The first category includes light charged hadrons, such as pions, kaons or protons, produced during the hadronisation process along with the signal \PB meson. It is referred to as a same-side-fragmentation (SSF) track category and is used by the SS taggers. The second category includes tracks associated with the decay of an opposite-side \Pb hadron, referred to as opposite-side-decay (OSD) tracks. These OSD tracks are the types of tracks used by the OS algorithms. Another category consists of light charged hadrons produced  in the fragmentation of the opposite-side $b$ hadron that are referred to as opposite-side-fragmentation (OSF)  tracks. The last category consists of the remaining tracks originating from the $\Pp\Pp$ collision, referred to as underlying tracks. These are considered background for the purposes of flavour tagging.

All four track categories are included in the IFT. The SSF, OSD and OSF categories carry information on the initial flavour of the signal \PB meson, in particular through the charge of the track and additional particle-identification (PID) information. The information provided by the OSD and OSF categories is nearly independent of the signal \PB-meson kinematics. On the other hand, the type of SSF track depends on the type of the light quark in the signal \PB meson. Pions and protons are produced along with a \Bz meson, while kaons are produced along with a \Bs meson. Therefore, to achieve an optimal performance, the classifiers for \Bz and \Bs mesons are trained independently on different samples. The \Bz tagger is trained on simulated events containing \mbox{$\Bz\to\jpsi\Kstarz$} decays and the \Bs tagger on simulated events containing \mbox{$\Bs\to\jpsi\phi$} decays.

Another aspect to consider for the construction of the neural network is the variation of the number of tracks in each event. The data samples of this study were collected at an instantaneous luminosity of $2\times 10^{33}\cm^{-2}\sec^{-1}$, for which the number of reconstructed tracks per event is approximately 40, with a small fraction of events containing more than 100 tracks. The DeepSets neural network architecture allows a variable number of inputs, making it a suitable choice of classifier. In addition, it provides an advantage in training duration that is an order of magnitude shorter than similar architectures such as recurrent neural networks. DeepSets architectures have been successfully employed in the development of flavour-tagging methods, such as in the OS jet tagger by the CMS collaboration~\cite{CMS:2024znt}.

\subsection{Training samples}
\label{sec:training_sel}
The simulated \mbox{$\Bz\to\jpsi\Kstarz$} and \mbox{$\Bs\to\jpsi\phi$} samples used for training are generated under Run~2 detector and accelerator conditions. Events must satisfy trigger requirements and additional selection criteria, consistent with those applied to data. The signal \PB-meson candidate is reconstructed from its charged decay products, while the remaining tracks in each event are used as input for the tagger training.

The chosen simulated samples correspond to decay channels with \PB-meson candidates selected by loose trigger requirements, ensuring coverage of a broad kinematic region.  
For the \( \Bz \) tagger, simulated  flavour-specific  \mbox{$\Bz\to\jpsi\Kstarz$} decays are used for training and the corresponding data sample, \mbox{$\Bz\to\jpsi\Kp\pim$}, is used for mistag calibration. The charge of the final-state pion identifies the flavour at decay time, enabling calibration through a time-dependent analysis of the flavour-tagged decay rate. 
For the \Bs tagger, training is performed using the \mbox{$\Bs\to\jpsi\phi$} simulated sample, while calibration of the mistag probability is done using a data sample of the flavour-specific  \mbox{$\Bs\to\Dsm\pip$} decay.  This strategy is chosen because a tagger trained on \mbox{$\Bs\to\jpsi\phi$} generalizes better to other \Bs decay modes, such as \mbox{$\Bs\to\Dsm\pip$}, than vice versa. 
This is primarily due to the broader kinematic range of \mbox{$\Bs\to\jpsi\phi$} decays enabled by less stringent trigger requirements, see Sec.~\ref{sec:detector}, which allows the classifier to learn across the full kinematic phase space. 

\subsection{The DeepSets classifier}
\label{sec:deepsets}
\begin{figure}[tb]
  \begin{center}
    \includegraphics[width=0.8\linewidth]{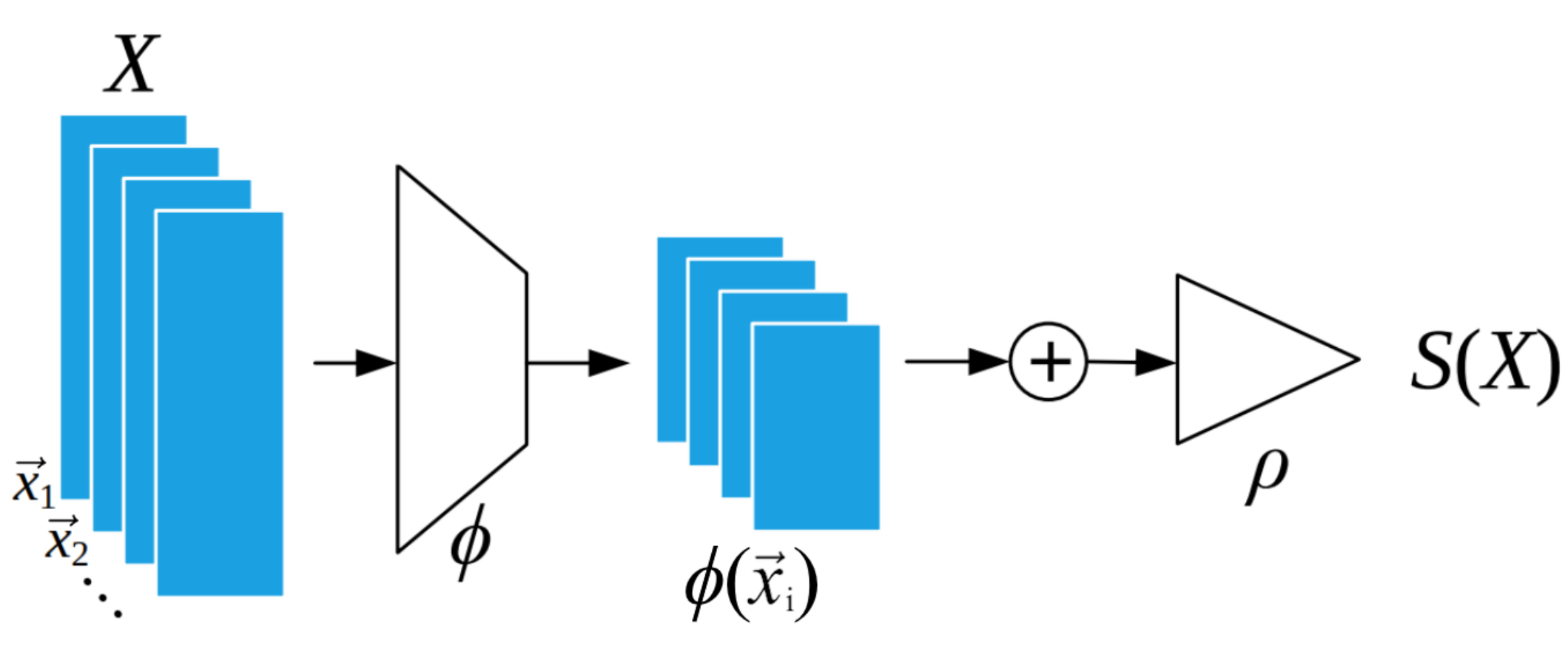}
    \vspace*{-0.5cm}
  \end{center}
  \caption{    
    Illustrative representation of DeepSets:
    The neural network response $S$ is inferred from the input matrix $X$ based on two processing steps. First, a feed-forward neural network, $\phi$, is applied to each feature vector $\vec{x_i}$ of the input matrix. The resulting sequence of output vectors $\phi(\vec{x_i})$ is then summed up into a single vector, which is evaluated by another feed-forward neural network, $\rho$. Figure adapted from Ref.~\cite{zaheer2018deep}.
    }
  \label{fig:DeepSet}
\end{figure}

The DeepSets~\cite{zaheer2018deep} process flow is illustrated in Fig.~\ref{fig:DeepSet}. Unlike a standard feed-forward network, which takes a single fixed-length feature vector as input, DeepSets operates on a variable-sized set of vectors $\vec{x}_i$.   Each input vector corresponds to a track and contains relevant features, which are described in the following sections. For convenience, these vectors are arranged in a matrix $X$. 
The matrix is passed through a feed-forward neural network, denoted $\phi$, which maps each input vector, \ie each matrix column, individually to a higher-level representation. The resulting set of transformed vectors is then aggregated through a permutation-invariant pooling operation (a simple summation). In this way, the overall output is insensitive to the ordering of the tracks, even though the $\phi$ network itself acts on individual tracks. Finally, the pooled representation is passed to a second feed-forward network, $\rho$, which produces the final network response, $S(X) = \rho(\sum_i \phi(\vec{x}_i))$. In the implementation used for the IFT, the $\phi$ network contains an input layer followed by two hidden layers of equal size, while the $\rho$ network consists of a hidden layer and an output layer with a single node.

The weights and biases of both networks are optimized during training. A rectified linear unit is used as the activation function in the hidden layers. The training is performed by minimizing the binary cross-entropy loss, evaluated over mini-batches and updated through backpropagation. A sigmoid activation ensures that the output of $S(X)$, the classification score $y$, is constrained to the interval $[0,1]$. The value of $y$ indicates whether the signal \PB meson contained a \Pb quark ($y=0$) or a \bquarkbar quark ($y=1$) at production.  In total, the network architecture contains 3890 trainable parameters. 

\subsection{Classifier training}
\label{sec:classifier_train}
The training of each tagger on its corresponding simulated sample proceeds in two steps.  First, to reduce the complexity of the task faced by the DeepSets network, a multiclass boosted decision tree~(BDT) classifier is trained using the XGBoost toolkit~\cite{Breiman,AdaBoost,Chen:2016:XST:2939672.2939785}.  
The BDT classifier assigns each track a probability of belonging to one of four categories: SSF, OSD, OSF and underlying event tracks. 
These category probabilities serve as inputs for the second stage, where a DeepSets neural network is trained to distinguish between the two \PB-meson production flavours.
The network's output is the final prediction score, denoted by $y$.

The tracks used for training are required to have hits in at least the VELO and the TT, and must additionally satisfy the following criteria: 
momentum between 2 and 200\gevc; ghost probability lower than 0.35; $\text{arctan}\left(\pt/p_z\right)>0.012$, where $p_z$ refers to the momentum component parallel to the beam axis, which ensures that the track exhibits a sufficient angular separation from the beam axis.
Both the BDT and the DeepSets classifiers use a common set of 23 track-related input features.
The feature selection builds on the set originally proposed in the paper introducing the IFT algorithm~\cite{likhomanenko}.
The features include:  
\begin{itemize}
    \item PID information on whether the track is an electron, a muon, a pion, a kaon or a proton; the probability that the track does not correspond to the genuine particle trajectory, referred to as ghost probability;

    \item track charge; \ptot and \pt of the track; component of the track momentum along the signal \PB-meson momentum direction;
 ionisation charge deposited in the silicon layers of the VELO by the track;
    \item IP of the track with respect to the decay vertex of the signal \PB-meson candidate; IP of a track with respect to the PV associated to the signal \PB meson and the difference in \chisq of a given PV reconstructed with and without the considered track; track-fit $\chi^2$;
    \item difference between the pseudorapidities of the track and the signal \PB meson; cosine of the angular difference in azimuth between the track and signal \PB-meson momentum directions; difference along the beam line direction between the track origin and the PV associated to the signal \PB meson;
    \item BDT-based variable that uses all underlying tracks in the event to determine the isolation of the track.
\end{itemize}
The DeepSets network, in addition, includes the four output probabilities of the track-classification BDT per track, as well as information from the five OS and three SS taggers, mentioned in Sec.~\ref{sec:introduction}. For each of these, the corresponding mistag probability and tagging decision are combined into a single input variable $d(1-2\eta)$. The addition of the SS and OS taggers leads to a relative improvement in tagging power of approximately 5\%. Note that even without them, the IFT performance exceeds that of the combined SS and OS taggers. 

Since the training set includes a substantial fraction of tracks with limited information about the \PB-meson production flavour, \eg tracks originating from different $pp$ collisions of the event, a study was carried out to reduce their impact. In events with multiple PVs, such tracks are likely to be classified as background by the BDT. 
To reduce their impact, a preselection based on the background-related BDT output was tested prior to applying the DeepSets classifier. While this preselection successfully removed about $20\%$ of the tracks without degrading tagging performance, it did not yield a significant improvement in training time or overall performance. Therefore, for simplicity, all tracks are retained in the current implementation.
Nevertheless, the study suggests potential avenues for reducing complexity in future developments, particularly relevant for the LHCb upgrades, where the number of PVs and associated tracks is expected to increase significantly.

\begin{figure}[tb]
  \begin{center}
    \hspace*{-1.cm}
    \includegraphics[width=0.49\linewidth]{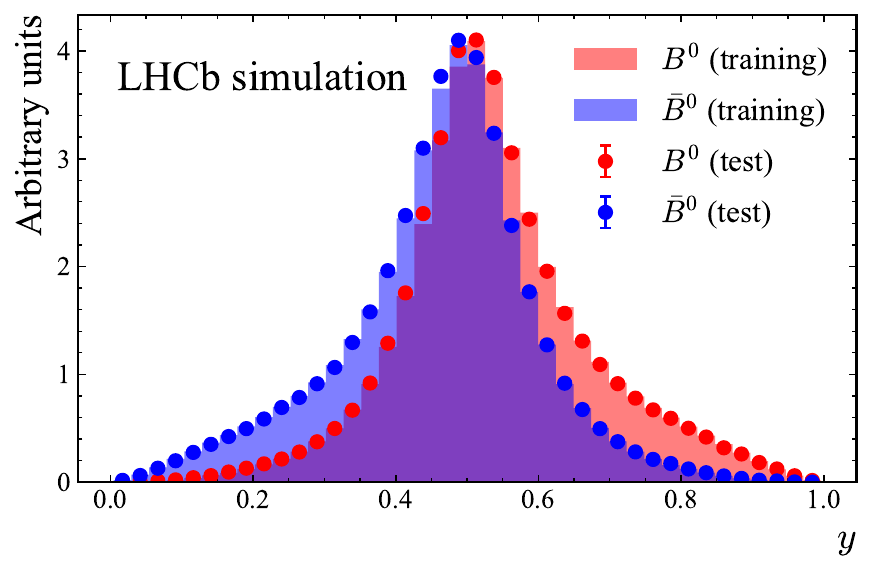}
    \includegraphics[width=0.49\linewidth]{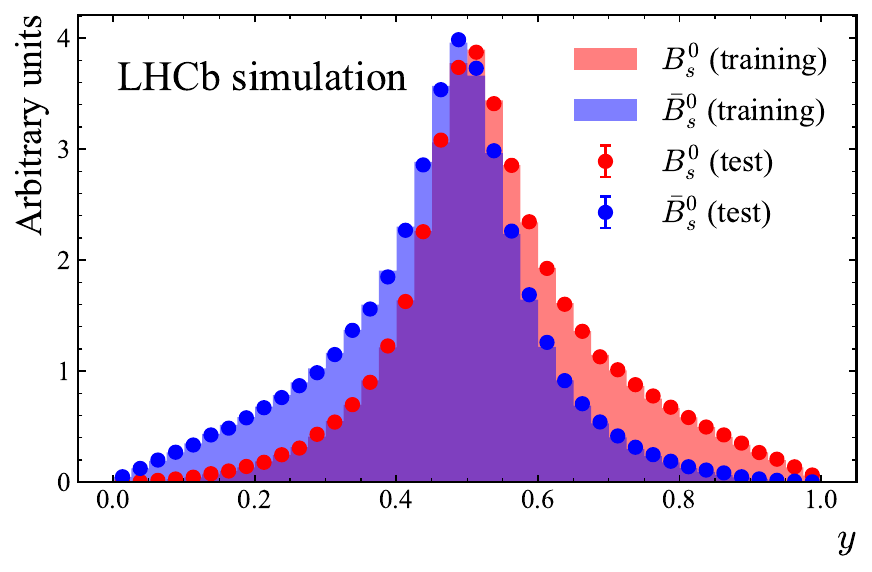}\\

    \includegraphics[width=0.49\linewidth]{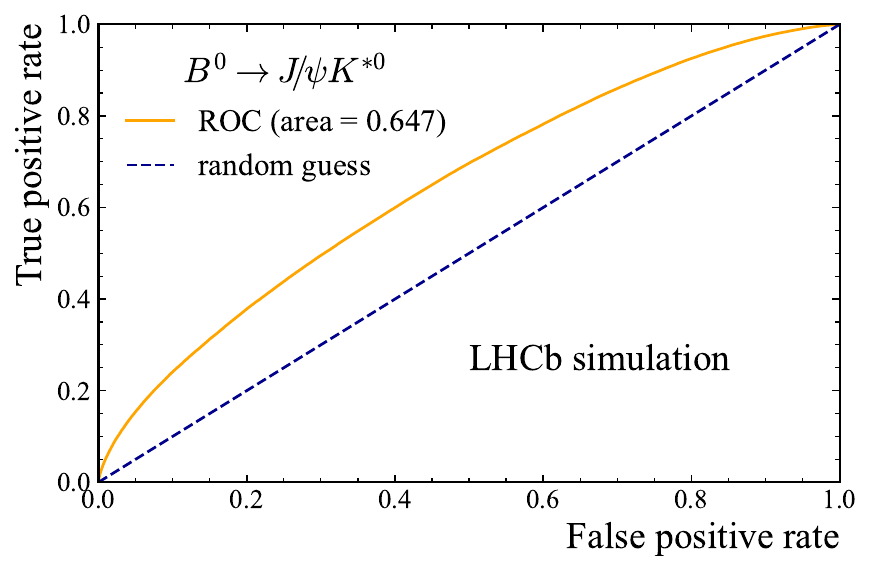}
   \includegraphics[width=0.49\linewidth]{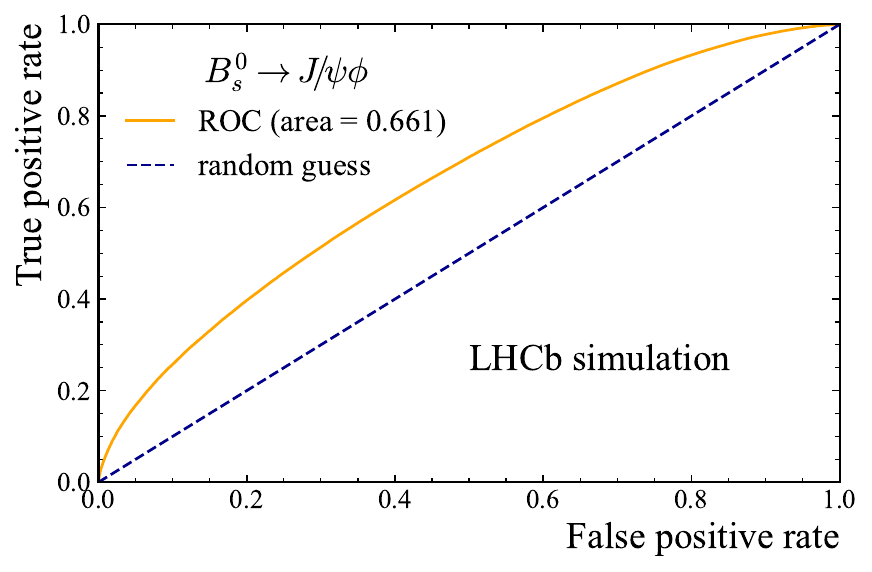}
    \vspace*{-0.5cm}
  \end{center}
  \caption{
    Distribution of the trained IFT classifier outputs $y$ obtained from (top left) \mbox{$\Bz\to \jpsi\Kstarz$}  and (top right) \mbox{$\Bs\to \jpsi \phi$}  simulation, showing the training subsets and the test subsets
    and ROC curve compared to a random guess function of the trained (bottom left) \mbox{$\Bz\to \jpsi \Kstarz$}  and (bottom right) \mbox{$\Bs\to \jpsi \phi$}  simulated samples. 
    }
  \label{fig:IFToutput}
\end{figure}

After the training, the distributions of the  classifier output for the \Bz and \Bs taggers are displayed in Fig.~\ref{fig:IFToutput}, showing a separation between the two \PB-meson flavours. In the same figure, the corresponding receiver operating characteristic (ROC) curves are shown. They display the rates of correctly tagged events (true positives) against those of mistagged events (false positives) and serve as an initial indicator of the performance of the classifier.

\section{Calibration of the mistag probability}
\label{sec:calibration}
The mistag probability estimated by the IFT algorithms is calibrated using  data samples of flavour-specific \mbox{$\Bz\to\jpsi\Kp\pim$} and \mbox{$\Bs\to \Dsm\pip$} decays. An unbinned maximum-likelihood fit to the decay-time $t$ is performed to resolve the neutral \PB-\Bbar flavour oscillation. The event-by-event decay-time resolution $\sigma_t$, estimated from a calibration of the decay-time uncertainty, is included as a conditional observable in the probability density function (PDF) $P(t|\sigma_t)$.
A precise understanding of the decay-time resolution is crucial, especially for $\Bs$ mesons, since a miscalibration can introduce a dilution of the oscillation amplitude that can be attributed to the flavour tagging.
The PDF is a convolution of the neutral \PB-meson time-dependent decay rate $\Gamma(t)$ and the decay-time resolution function $G(t|\sigma_t)$, multiplied by the decay-time acceptance function $\epsilon(t)$,
\begin{equation}
 P(t|\sigma_t) \propto \epsilon (t) \left[ \Gamma(t')\ast G(t - t' | \sigma_t) \right].
 \label{eq:timeacc}
\end{equation} 
The decay-time acceptance for both calibration channels is modelled by cubic splines, whose parameters are allowed to vary in the fit. The decay-time resolution is modelled by a single Gaussian function, where $\sigma_t$ is a per-candidate conditional variable. The neutral-\PB-meson time-dependent decay rate, 
\begin{equation}
\label{eq:dt-fit}
    \Gamma(t| \eta, C_\text{phys}) \propto e^{-\Gamma_{d,s} t} 
    \left[ \left(1 - C_\text{phys} \Delta \omega(\eta) \right) \cosh \left( \frac{1}{2} \Delta \Gamma_{d,s} t \right) + C_\text{phys} \left( 1 - 2 \omega(\eta)\right) \cos \left( \Delta m_{d,s} t \right) \right] ,
\end{equation}
is a function of the difference between the heavy and light decay widths, $\Delta\Gamma_{\Pd,\Ps}$, and mass differences $\Delta m_{\Pd,\Ps}$, where  the subscript \Pd denotes \Bz mesons and \Ps denotes \Bs mesons.
The value of $C_{\text{phys}}$ is $+1$ if the \PB-meson flavour at decay, identified by the charge of the final-state pion, matches its flavour at the time of production, determined from the tagging decision. 
If the flavours do not match, $C_{\text{phys}}$ is set to $-1$, and if the candidate is untagged it is set to $0$.
The parameters $\Delta\Gamma_s$, $\Delta m_{d,s}$, and $\Gamma_{d,s}$ are constrained to their known values by applying Gaussian penalty terms based on their respective uncertainties~\cite{LHCb-PAPER-2023-016,PDG2024,HFLAV23}.
 The value of $\Delta\Gamma_d$ is fixed to zero.  The calibrated mistag probability $\omega$ is parametrised as a linear function of the estimated probability $\eta$,
\begin{equation}
\begin{aligned}
    \omega^{\PB}(\eta) &= p_0 + \frac{1}{2}  \Delta p_0 + \left(p_1 + \frac{1}{2} \Delta p_1\right) (\eta - \bar{\eta}), \\
    \omega^{\Bbar}(\eta) &= p_0 - \frac{1}{2}  \Delta p_0 + \left(p_1 - \frac{1}{2} \Delta p_1\right) (\eta - \bar{\eta}),
    \label{eq:omegavseta}
\end{aligned}
\end{equation}
where $p_0$ and  $p_1$ are the linear function offset and slope, and $\bar{\eta}$ is the average estimated mistag probability of the sample. The factor \mbox{$\Delta \omega(\eta) = \Delta p_0+\Delta p_1(\eta-\bar{\eta})$} accounts for different mistag probabilities between \PB and \Bbar mesons. A second-order polynomial fit yields coefficients consistent with those of a linear model and therefore, the simpler functional form is adopted. 
Events in which the calibrated mistag probability exceeds 0.5 are treated as untagged.
This results in a calibrated tagging efficiency lower than $100\%$ for the IFT.

The ideal calibration value $p_{1}=1$ corresponds to the case where the predicted mistag $\eta$ equals the calibrated mistag probability $\omega$ and the linear calibration model is exact. The values  $\Delta p_{0,1} = 0$ indicate that there is no difference between the tagging of \PB mesons and \Bbar mesons. In practice, deviations from the ideal case arise from differences between data and simulation, residual model bias, detection asymmetries, model nonlinearity and species-dependent effects. This is what causes that $\omega$ differs from $\eta$, which is obtained from training on simulation. The calibration on data aims at correcting such discrepancies by matching $\omega$ to the observed mistag fraction in the data. Such effects have been observed in tagging algorithms both within and beyond LHCb (see e.g. Refs.~\cite{LHCb-PAPER-2015-056, LHCb-PAPER-2016-039, Belle-II:2024lwr}).

\subsection{\boldmath Calibration of the \Bz classifier}
\label{sec:Bd_cal}
To reduce background contributions, loose requirements are applied on \mbox{$\Bz\to \jpsi \Kp\pim$} candidates, based on Ref.~\cite{LHCb-PAPER-2023-016}.
These include \pt and PID selections, which reduce backgrounds caused by misidentifying the pion as either a kaon or a proton. 

The dominant remaining background is combinatorial, with a minor contribution from \mbox{$\Bs \to \jpsi\Kp\pim$} decays. 
These background contributions are subtracted using the \sPlot~\cite{Pivk:2004ty} technique, based on an unbinned maximum-likelihood fit to the invariant mass of the \PB-meson candidates. 

\begin{figure}[tb]
\begin{center}
\includegraphics[width=0.6\linewidth]
{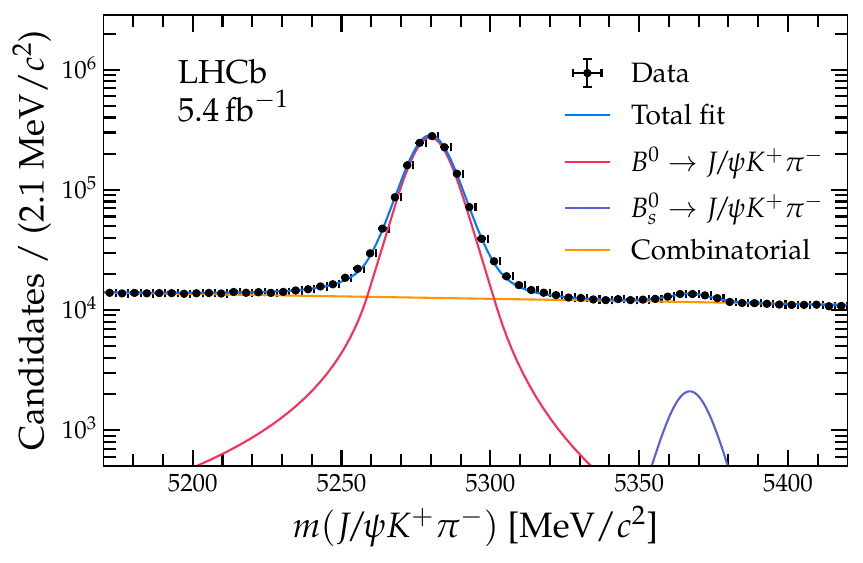}
\end{center}
\caption{
  Invariant-mass distribution of  \mbox{$\Bz \to \jpsi\Kp\pim$} candidates in the combined data sample with the result of the fit also shown.
}
\label{fig:mass_fit_Bz}
\end{figure}

The mass distribution and corresponding fit result for the data sample combined over all data-taking years and $C_\text{phys}$ values are shown in Fig.~\ref{fig:mass_fit_Bz}.
The signal mass distribution is modelled with a Hypatia function~\cite{Santos:2013gra} where the tail parameters are fixed from simulation, while the mean and width of the Gaussian core are free in the fit to the data. The same functional form is used to describe the \mbox{$\Bs\to \jpsi\Kp\pim$} contribution, differing only by a shift in the mean corresponding to the known $\Bs-\Bz$ mass difference ~\cite{PDG2024}.
An exponential function is used to describe the combinatorial background contribution and its parameters are free in the fit to the data.
The resulting signal yields are $399\times 10^3$, $394\times 10^3$ and $471\times 10^3$ for 2016, 2017 and 2018, respectively.

The \Bz meson has an oscillation period that is significantly larger than the \lhcb detector decay-time resolution. 
Therefore, for simplicity, the calibration of the decay-time resolution of \mbox{$\Bz\to \jpsi \Kp\pim$} decays is obtained from simulated decays instead of data. 
For this, the simulated dataset is split into eight bins of the reconstructed decay time. 
For each bin $i$ an unbinned maximum-likelihood fit to the difference of the reconstructed and the true decay time is performed and is taken as the time resolution $\sigma_{t,i}$ of this bin.
Subsequently, the effective decay-time resolution is obtained through a linear fit on $\sigma_{t,i}$ values.
A linear dependence between the estimated and calibrated decay-time resolutions is assumed. To account for possible differences with the decay-time resolution in data, it is verified that variations of 5 times the statistical uncertainties on the calibration parameters do not affect the flavour-tagging performance. 
These variations also cover the case where no calibration is applied (\ie, the calibrated decay-time resolution is identical to the uncalibrated one). This demonstrates that the decay-time resolution for \Bz decays has little influence on the tagging calibration.

\begin{figure}[tb]
\begin{center}
\includegraphics[width=0.6\linewidth]{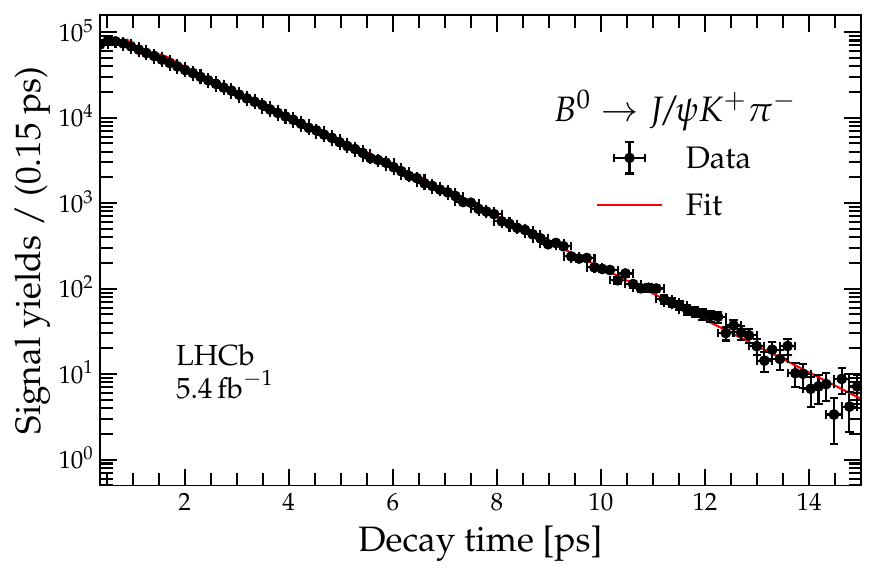}
\end{center}
\caption{
Background-subtracted decay-time distribution of \mbox{$\Bz \to \jpsi\Kp\pim$} candidates in the data sample  combined across all data-taking years and initial and final flavours, together with the fit result.
}
\label{fig:bd_IFT_timefit}
\end{figure}

\begin{figure}[tb]
\begin{center}
\includegraphics[width=0.6\linewidth]{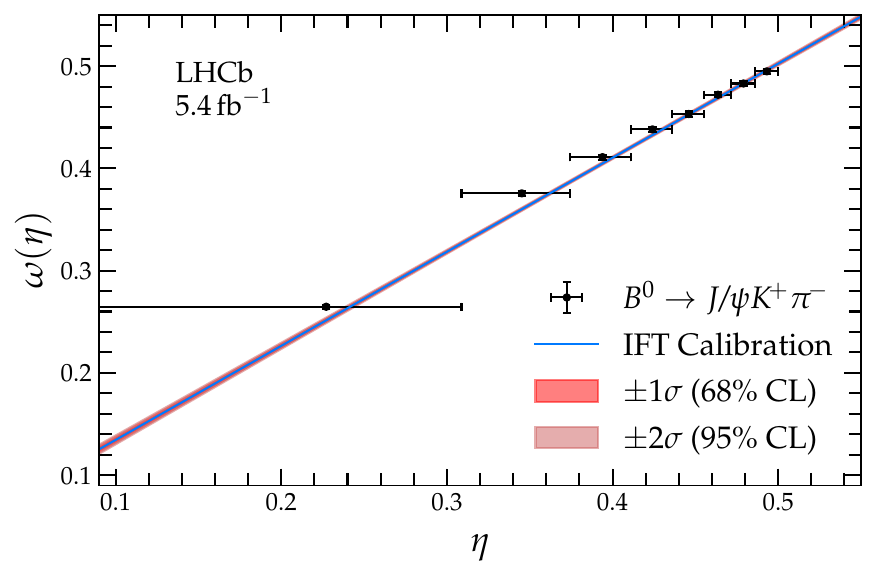}
\end{center}
\caption{
   Measured mistag fraction $\omega$ as a function of the mistag probability $\eta$ in $\Bz\to\jpsi\Kp\pim$ data. The data points are obtained by performing a decay-time fit in each $\eta$ bin and extracting the fraction of mistagged events.
}
\label{fig:bd_IFT_cal}
\end{figure}

A maximum-likelihood fit to the background-subtracted decay time of the \mbox{$\Bz\to \jpsi \Kp\pim$} sample is performed using Eq.~\ref{eq:timeacc} with inputs from Eq.~\ref{eq:dt-fit}. 
The fit result combined across all data-taking years and initial and final flavours is shown in Fig.~\ref{fig:bd_IFT_timefit}. The resulting calibration function is shown in Fig.~\ref{fig:bd_IFT_cal}. 
The calibration parameters are summarised in Table~\ref{tab:cal_bd}. Their mutual correlations do not exceed 10\%. Differences in the calibration parameters between data-taking years are small, yet significant enough to justify determining them independently.

\begin{table}
\centering
\caption{Results of the unbinned decay-time fit for the IFT calibration parameters in \mbox{$\Bz \to \jpsi \Kp\pim$} data as well as the average uncalibrated mistag estimation. The uncertainties are statistical.}
\resizebox{1\columnwidth}{!}{
\begin{tabular}{lccccc}
\toprule
Year      & $p_0            $ & $p_1            $ & $\Delta p_0     $ & $\Delta p_1     $ & $\bar{\eta}$\\  
\midrule
2016                             & $0.4146 \pm 0.0012              $ & $0.896 \pm 0.012                $ & $\phantom{+}0.0006 \pm 0.0017              $ & $-0.189 \pm 0.018               $ & $0.4056$\\ 
2017                             & $0.4124 \pm 0.0012              $ & $0.922 \pm 0.012                $ & $-0.0016 \pm 0.0017             $ & $-0.234 \pm 0.018               $ & $0.4044$\\ 
2018                             & $0.4128 \pm 0.0011              $ & $0.926 \pm 0.011                $ & $-0.0021 \pm 0.0016             $ & $-0.182 \pm 0.016               $ & $0.4038$\\ 
\bottomrule
\end{tabular}
}
\label{tab:cal_bd}
\end{table}

\subsection{\boldmath Calibration of the \Bs classifier}
\label{sec:Bs_cal}

The $\Bs \to \Dsm\pip$ candidates are selected through the $\Dsm \to \Kp\Km\pim$ decay mode following the strategy of Ref.~\cite{LHCb-PAPER-2021-005}.  The purity of the sample is improved by imposing PID requirements on pions and kaons. Additional vetoes on the invariant masses of the \Dsm decay products are applied to suppress backgrounds from \Dz decays and from misidentified particles originating from \Dm and $\PLambda_c^-$ decays.

\begin{figure}[tb]
\begin{center}
\includegraphics[width=0.6\linewidth]
{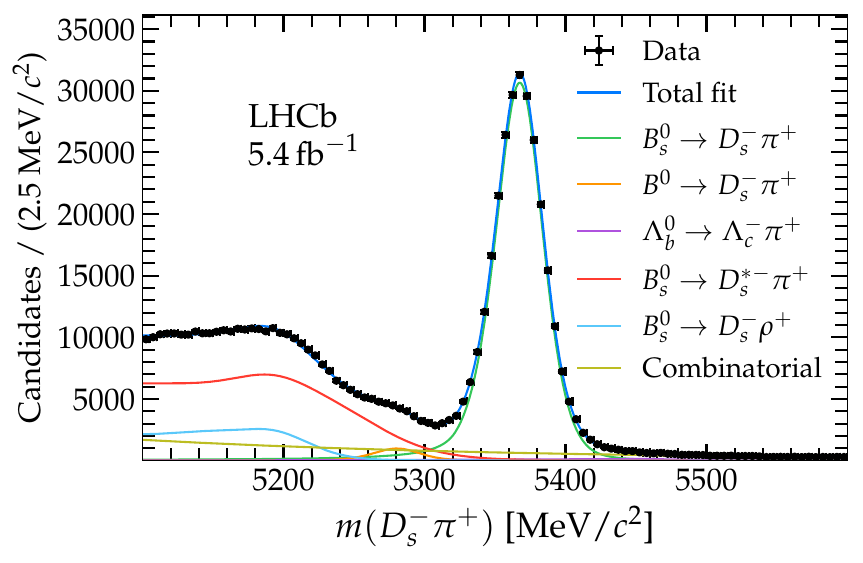}
\end{center}
\caption{
  Invariant-mass distribution of  \mbox{$\Bs\to\Dsm\pip$} candidates in the combined data sample with the result of the fit also shown. }
\label{fig:mass_fit_bs}
\end{figure}

As in the case of the \Bz classifier, the background in data is subtracted using the \sPlot technique.  The distribution of the \Bs-meson candidate mass along with the fit to the data is shown in Fig.~\ref{fig:mass_fit_bs}. Following the strategy of Ref.~\cite{LHCb-PAPER-2023-016}, the signal mass model is described by a Hypatia function and the combinatorial background is modelled with an exponential function. The background shapes of \Pb-hadron decays, including $\Bs\to \Dsm\Kp$, $\Bz \to \Dm \pip$, $\Bs \to D_s^{* -}\pi^+$, \mbox{$\Bs \to \Dsm \rho^+$} and $\PLambda_b^0 \to \PLambda_c^- \pip$, are modelled based on simulated samples. The background subtraction is performed in two steps. First, a fit to the invariant-mass distribution of the \Bs candidates in the region $[5100,5600]\mevcc$ is performed, in which the signal shape parameters are fixed to the values obtained from simulation. The shape parameter of the combinatorial background and the fractions of partially reconstructed and misidentified $b$-hadron decays are free to vary in the fit. In the second step, a fit is performed in the narrower mass window $[5300,5600]\mevcc$ to extract the {\it sWeights}. Here, all parameters are fixed to the results of the first fit, except for the signal and total background yields, which are allowed to vary.
 The resulting $\Bs \to \Dsm\pip$ signal yields are $85\times 10^3$, $87\times 10^3$ and $101\times 10^3$ for 2016, 2017 and 2018, respectively.

In the fit to the decay-time distribution, the decay-time acceptance is modelled, as in the case of $\Bz\to\jpsi\Kp\pim$ decays, with cubic splines. The cubic spline parameters are free in the fit to the data. 

\begin{figure}[tb]
\begin{center}
\includegraphics[width=0.6\linewidth]{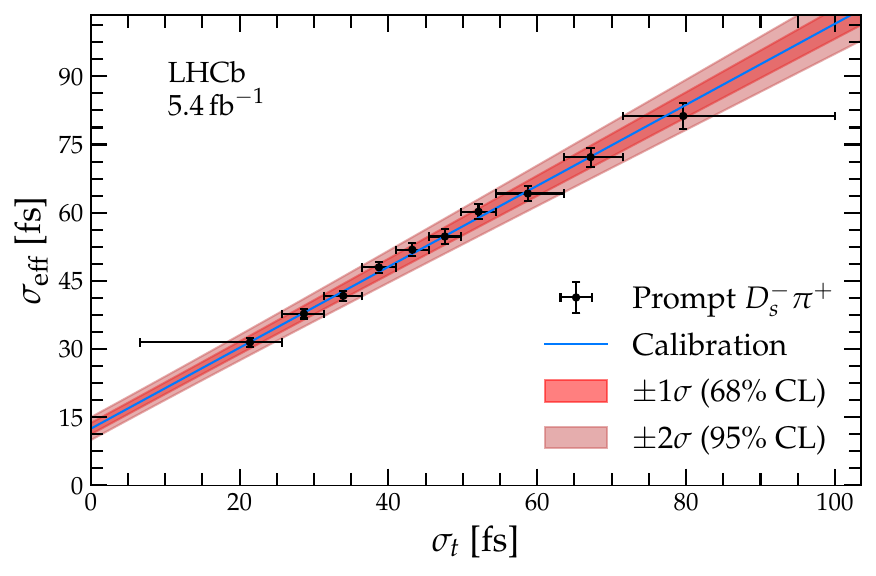}
\end{center}
\caption{
   Measured effective decay-time resolution $\sigma_\text{eff}$ as a function of the reconstructed decay-time resolution $\sigma_t$ in prompt \mbox{$\Dsm \pip$} data. 
}

\label{fig:bs_timeres_cal}
\end{figure}

The \Bs decay-time resolution, $\sigma_t$, is calibrated in a fit to the decay-time distribution of background-subtracted prompt $\Dsm\to\Kp\Km\pim$ decays produced from $\Pp\Pp$ collisions combined with an opposite-charge track. These are selected in the same way as the $\Bs \to \Dsm\pip$ decays, except for the requirement on the \Bs candidate decay time, which is not imposed.
An unbinned maximum-likelihood fit to the decay-time distribution of the candidates  is performed in each of ten $\sigma_t$ bins using a PDF proportional to
\begin{equation}
\label{eq:sigma_eff} f_\text{sig}\cdot\left[(f_\text{prompt}\delta(t) + f_{ll}e^{-t/\tau_2})*G(t-t') \right] + f_\text{wpv}e^{-|t|/\tau_1},
\end{equation}
where the resolution function $G(t)$ consists of the sum of two Gaussian functions with a common mean, two standard deviations  $\sigma_1$ and $\sigma_2$ and a relative fraction $f_1$~\cite{LHCb-PAPER-2019-013}. The prompt signal component, contributing with a fraction $f_\text{prompt}$, is assumed to have zero lifetime and is modelled with a Dirac $\delta$ distribution.
 In addition, two components with finite lifetimes, $\tau_1$ and $\tau_2$, and fractions $f_\text{wpv} \equiv(1-f_\text{sig})$
 and $f_{ll}\equiv(1-f_\text{prompt})$ are modelled with exponential functions. The exponential functions represent prompt \Dsm candidates associated to a wrong PV, and \Dsm mesons originating from decays of heavier particles.
The standard deviations of the two Gaussian components in $G(t-t')$ are combined into a single effective resolution, $\sigma_\text{eff}$, which yields the same dilution of the \Bs-meson oscillation amplitude, $D_t$, as the resolution model $G(t-t')$ in Eq.~\ref{eq:sigma_eff}.
This effective resolution is computed according to
\begin{equation}
\begin{split}
    D_{t} &= f_1 e^{-\sigma_1^2\Delta m_s^2/2} + (1-f_1) e^{-\sigma_2^2\Delta m_s^2/2} , \\
    \sigma_\text{eff} &= \sqrt{-2 \ln{D_{t}}}/\Delta m_s .
\end{split}
\end{equation}
Differences in the selection between the prompt $\Dsm \pip$ calibration sample and the \mbox{$\Bs\to\Dsm\pip$} sample are accounted for by scaling $\sigma_t$ according to differences among the corresponding simulated samples. 
A $\chi^2$ fit is performed in bins of the  measured decay-time resolution $\sigma_t$  assuming a linear dependency of $\sigma_\text{eff}$ on $\sigma_t$. The results are shown in Fig.~\ref{fig:bs_timeres_cal}.
The results of this fit are used as a per-candidate conditional variable, which are converted to the standard deviation of the Gaussian decay-time resolution model in Eq.~\ref{eq:timeacc}.

\begin{figure}[tb]
\begin{center}
\includegraphics[width=0.6\linewidth]{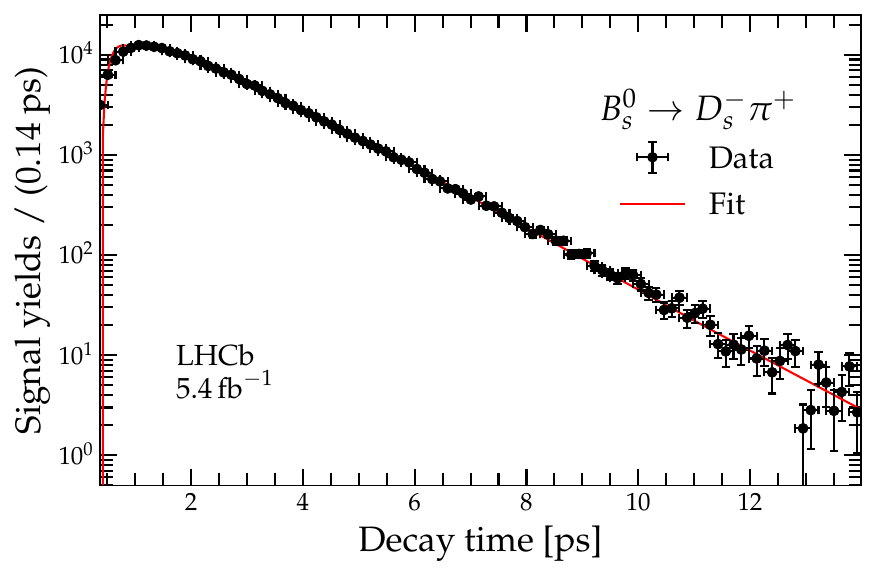}
\end{center}
\caption{
   Background-subtracted decay-time distribution of \mbox{$\Bs\to\Dsm\pip$} candidates in the data sample  combined across all data-taking years and initial and final flavours, together with the fit result.
}
\label{fig:bs_IFT_timefit}
\end{figure}

\begin{figure}[tb]
\begin{center}
\includegraphics[width=0.60\linewidth]{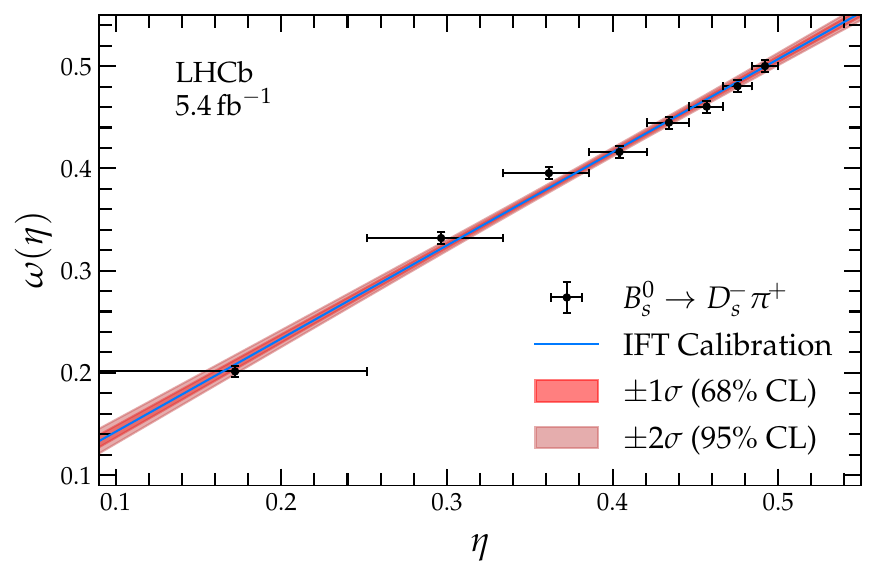}
\end{center}
\caption{
   Distribution of 
 the measured mistag fraction $\omega$ as a function of the mistag probability $\eta$ in $\Bs\to\Dsm \pip$ data. The data points are obtained by performing a decay-time fit in each $\eta$ bin and extracting the fraction of mistagged events.}
\label{fig:bs_IFT_cal}
\end{figure}

A maximum-likelihood fit to the background-subtracted decay time of the \mbox{$\Bs\to \Dsm\pip$} sample is performed. The fit results are shown in Fig.~\ref{fig:bs_IFT_timefit}. The resulting IFT calibration function is shown in Fig.~\ref{fig:bs_IFT_cal}. The calibration parameters are summarised in Table~\ref{tab:cal_bs}. The correlations between them do not exceed 5\%.

\begin{table}[tb]
\caption{Results of the unbinned decay-time fit for the IFT calibration parameters in \mbox{$\Bs \to \Dsm \pip$} data as well as the average uncalibrated mistag estimation. The uncertainties are statistical.}
\centering
\resizebox{1\columnwidth}{!}{
\begin{tabular}{lccccc}
\toprule
Year                      & $p_0            $ & $p_1            $ & $\Delta p_0     $ & $\Delta p_1     $ & $\bar{\eta}$\\ 
\midrule
2016                            & $0.4038 \pm 0.0041              $ & $0.954 \pm 0.037                $ & $0.0147 \pm 0.0037              $ & $-0.083 \pm 0.035                $ & $0.3872$\\        
2017                            & $0.4014 \pm 0.0036              $ & $0.877 \pm 0.032                $ & $0.0062 \pm 0.0036             $ & $\phantom{+}0.027 \pm 0.034                $ & $0.3871$ \\ 
2018                            & $0.4022 \pm 0.0033              $ & $0.880 \pm 0.030                $ & $0.0123 \pm 0.0034              $ & $-0.046 \pm 0.031                $ & $0.3864$ \\ 
\bottomrule
\end{tabular}
}
\label{tab:cal_bs}
\end{table}

\section{Performance}
\label{sec:performance}

The performance of the IFT algorithm after calibration is evaluated on data and compared to the performance achieved by combining the existing \lhcb flavour-tagging algorithms, the OS and the SS taggers. The five OS taggers are used to compare both the \Bz and the \Bs taggers. On the other hand, given the different hadronisation processes for \Bz and \Bs mesons, the SS taggers used in the comparison are based on pions and protons for \Bz and on kaons for \Bs decays.

The combination of multiple tag decisions, $d_i$, and mistag estimations, $\eta_i$, is obtained by calculating the probability of the signal candidate to contain a \Pb or a $\bquarkbar$ quark, denoted as $P(\Pb)$ and $P(\bquarkbar)$,
\begin{equation}
    P(\Pb) = \frac{p(\Pb)}{p(\Pb)+p(\bquarkbar)}, \hspace{0.2cm} P(\bquarkbar) = 1 - P(\Pb) ,
\end{equation}
where the probabilities $p(b)$ and $p(\bquarkbar)$ for uncorrelated tagging decisions are defined as
\begin{equation}
\begin{split}
    p(b) &= \prod_i \left( \frac{1 + d_i}{2} - d_i (1 - \eta_i) \right), \\
    p(\bquarkbar) &= \prod_i \left( \frac{1 - d_i}{2} + d_i (1 - \eta_i)\right).
\end{split}
\end{equation}
The combined tagging decision and the combined mistag estimate are given by
\begin{equation}
\begin{split}
    d_\text{comb} &= \text{sign}(P(\bquarkbar) - P(b)), \\
    \eta_\text{comb} &= 1 - \text{max}(P(b), P(\bquarkbar) ).
\end{split}
\end{equation}

The combination of the SS and OS taggers proceeds in three steps. First, the individual OS taggers are combined into a single OS decision and mistag estimate. Independently of the OS taggers, the two SS taggers used in the \Bz classifier are combined into a single SS decision and mistag estimate. 
Subsequently, the OS combination mistag estimate and, in the case of the \Bz classifier, the SS combination or, in the case of the \Bs classifier, the SS kaon mistag estimate, are calibrated according to Eqs.~\ref{eq:omegavseta}. In the final step, the calibrated OS combination and the SS combination or SS kaon are again combined into a single decision and mistag estimation, which is further calibrated according to Eqs.~\ref{eq:omegavseta}.
Performing the calibration after combining the taggers is essential to properly account for potential correlations between them, which could otherwise bias the resulting mistag.
The calibrations are verified to be in agreement with those obtained using a dedicated flavour-tagging calibration software package {\tt lhcb-ftcalib}~\cite{lhcb_ftcalib}.

\subsection{\boldmath Performance of the \Bz classifier}
The performance of the \Bz classifier on data is determined for two decay modes, \mbox{$\Bz\to\jpsi\Kp\pim$} and \mbox{$\Bz\to\jpsi\KS$} decays. It is compared to the performances of the SS and OS taggers and their combination. The performances on the \mbox{$\Bz\to\jpsi\Kp\pim$} samples are summarised in Table~\ref{tab:tag_power_bd_comparison}. A relative tagging power increase of at least 35\% over the combination of SS and OS taggers is observed.
\begin{table}
\caption{Performances on \mbox{$\Bz\to\jpsi\Kp\pim$} data for the combination of SS and OS taggers versus the IFT. The uncertainties are statistical.  }
\label{tab:tag_power_bd_comparison}
\centering
\begin{tabular}{llcccc}
\toprule
              Year & Tagger   & $\varepsilon_\text{tag} (\%)$                                    & $\langle D^2\rangle$                          & $\varepsilon_\text{tag} \langle D^2\rangle (\%)$              \\
\midrule

\multirow{2}{*}{2016}  & SS and OS &$87.030 \pm 0.058                                  $&$0.0466 \pm 0.0010             $&$4.05 \pm 0.09               $\\
 & $\mathrm{IFT}                            $&$95.789 \pm 0.035                                  $&$0.0582 \pm 0.0011             $&$5.58 \pm 0.10                 $\\
\multirow{2}{*}{2017} & SS and OS &$87.208 \pm 0.058                                  $&$0.0485 \pm 0.0010             $&$4.23 \pm 0.09              $\\
 & $\mathrm{IFT}                            $&$95.024 \pm 0.038                                  $&$0.0617 \pm 0.0011             $&$5.87 \pm 0.10                 $\\
\multirow{2}{*}{2018} & SS and OS &$87.631 \pm 0.052                                  $&$0.0493 \pm 0.0010           $&$4.32 \pm 0.09               $\\
 & $\mathrm{IFT}                            $&$95.172 \pm 0.034                                  $&$0.0615 \pm 0.0010             $&$5.85 \pm 0.10               $\\

\bottomrule
\end{tabular}
\end{table}
\begin{table}
\caption{Performance of the IFT on \mbox{$\Bz\to\jpsi\Kp\pim$} data for all years, evaluated separately for events with only SS tagger decisions, only OS tagger decisions, both SS and OS tagger decisions, at least one tagger decision (SS or OS), and no tagger decision from either SS or OS taggers.}
\label{tab:tag_power_bd_splits}
\centering
\begin{tabular}{lccc}
\toprule
      Configuration          & $\varepsilon_\text{tag} (\%)$                                    & $\langle D^2\rangle$                          & $\varepsilon_\text{tag} \langle D^2\rangle (\%)$              \\
\midrule
    only SS & $51.79 \pm 0.05$ & $0.0343 \pm 0.0008$ & $1.77 \pm 0.03 $\\
    only OS & $\phantom{0}4.90 \pm 0.02$ & $0.1239 \pm 0.0031$ & $0.61 \pm 0.02 $\\
    SS and OS & $28.75 \pm 0.05$ & $0.1126 \pm 0.0013$ & $3.24 \pm 0.04 $\\
    SS or OS & $85.06 \pm 0.03$ & $0.0664 \pm 0.0006$ & $5.64 \pm 0.06 $\\
    neither & $10.93 \pm 0.02$ & $0.0128 \pm 0.0020$ & $0.14 \pm 0.01 $\\

\bottomrule
\end{tabular}
\end{table}

The IFT tagging power on \mbox{$\Bz\to\jpsi\Kp\pim$} data in relation to the tagging decision of the SS and OS taggers across all years is shown in Table~\ref{tab:tag_power_bd_splits}. 
For this study, the dataset is divided into five subsets based on the tagging decisions: events with only an SS tagger decision, only an OS tagger decision, both SS and OS tagger decisions, at least one tagger decision (SS or OS), and events with neither SS nor OS tagger decisions. A recalibration is then performed separately on each of these subsets using the full dataset spanning all years. 
Since events with $\omega>0.5$ are treated as untagged, recalibrating each subset independently results in total efficiencies and performances that differ from those computed on the full dataset.
The absolute tagging power obtained in events not tagged by either the SS or OS taggers represents only a small fraction of the overall improvement in tagging power achieved by the IFT. In contrast, the IFT tagging power for events tagged either by the SS or the OS taggers, accounts for the dominant share of the improvement with respect to using the SS and OS taggers alone. 
This indicates that the improvement in the tagging power is driven by additional information the IFT uses in events tagged by SS and OS.

The \Bz IFT was developed using simulated \mbox{$\Bz\to\jpsi\Kstarz$} decays. To assess whether its performance improvement over the current flavour-tagging algorithm is also observed in other decay modes, the calibration from the control channel is adapted to the \mbox{$\Bz\to\jpsi\KS$} decay in data, and the tagging power is subsequently measured. 
Decays of \mbox{\decay{\KS}{\pip\pim}} are reconstructed from two tracks forming a
good-quality vertex in two different categories: the first one consists of \KS mesons that decay within the acceptance of the vertex detector, and the second contains \KS that decay outside of the vertex detector. 
The selection criteria are based on the latest measurement of $\text{sin}(2\beta)$~\cite{LHCb-PAPER-2023-013}, though they are simplified by not applying a full multivariate selection.
 The resulting signal yields are $29\times 10^3$, $30\times 10^3$ and $34\times 10^3$ for 2016, 2017 and 2018, respectively.

The mistag probability for all taggers is calibrated using $\Bz\to\jpsi\Kp\pim$ decays which are weighted so that the distributions of the \pt and pseudorapidity of the signal \PB meson, the number of tracks and the number of PVs match those in \mbox{$\Bz \to \jpsi \KS$} data. 
\begin{table}
\caption{Performances evaluated for \mbox{$\Bz\to \jpsi \KS$} data for the combination of SS and OS taggers and the IFT. The uncertainties are statistical. }
\label{tab:tag_power_JpsiKS}
\centering
\begin{tabular}{llccc}
\toprule
      Year & Tagger           & $\varepsilon_\text{tag} (\%)$                                    & $\langle D^2\rangle$                          & $\varepsilon_\text{tag} \langle D^2\rangle (\%)$              \\
\midrule
\multirow{2}{*}{2016}  &  SS and OS & $69.18 \pm 0.35$ & $0.0473 \pm 0.0011$ & $3.27 \pm 0.08                 $\\
& $\mathrm{IFT}                            $&$95.42 \pm 0.16                                  $&$0.0563 \pm 0.0012             $&$5.37 \pm 0.12                 $\\
\multirow{2}{*}{2017} &  SS and OS &$68.65 \pm 0.34                                   $&$0.0482 \pm 0.0011             $&$3.31 \pm 0.08                 $\\
& $\mathrm{IFT}                            $&$98.53 \pm 0.09                                  $&$0.0571 \pm 0.0012            $&$5.63 \pm 0.12                 $\\
\multirow{2}{*}{2018} & SS and OS                            &$69.58 \pm 0.31                                   $&$0.0487 \pm 0.0010             $&$3.38 \pm 0.07                $\\
& $\mathrm{IFT}                            $&$96.41 \pm 0.13                                  $&$0.0573 \pm 0.0011            $&$5.52 \pm 0.11                 $\\
\bottomrule
\end{tabular}
\end{table}
Table~\ref{tab:tag_power_JpsiKS} compares the calibrated tagging power across all years for the SS and OS tagger combination and the IFT on \mbox{$\Bz\rightarrow \jpsi\KS$} data. Interestingly, the relative improvement over the SS and OS taggers is greater than in \mbox{$\Bz \to \jpsi\Kp\pim$} data. This is due to differences in tagging efficiency, as shown in Tables~\ref{tab:tag_power_bd_comparison} and~\ref{tab:tag_power_JpsiKS}.
In the recent \mbox{$\sin(2\beta)$} measurement, which uses a more refined candidate selection, the tagging efficiency is $(85.34 \pm 0.05)\%$, similar to \mbox{$\Bz \to \jpsi\Kp\pim$} decays, see Table~\ref{tab:tag_power_bd_comparison}. Differences in the tagging dilution between this study and the \mbox{$\sin(2\beta)$} measurement are negligible. Therefore, the reduced tagging power of the SS and OS taggers compared to that measurement is attributed to lower tagging efficiency resulting from the simplified signal selection. In contrast, the performance of the IFT remains robust under these changes. By correcting the efficiency to one of the \mbox{$\Bz\to \jpsi\KS$}  samples selected as in the \mbox{$\sin(2\beta)$} measurement to $85\%$, while keeping the dilution constant, the IFT improvement aligns with that observed in $\Bz\to\jpsi\Kp\pim$ decays.

\subsection{\boldmath Performance of the \Bs classifier}
The tagging performance of the \Bs classifier is evaluated on \mbox{$\Bs\to\Dsm\pip$} and \mbox{$\Bs\to\jpsi\Kp\Km$} data using the calibration from \mbox{$\Bs\to\Dsm\pip$} data. 
The selection is identical to that developed for the $\phi_s$ measurement~\cite{LHCb-PAPER-2023-016}. 
The resulting signal yields are $99\times 10^3$, $103\times 10^3$ and $123\times 10^3$ for 2016, 2017 and 2018, respectively. 

An improvement of the IFT over the SS and OS combination at the level of approximately $20\%$ is observed, as shown in Tables~\ref{tab:tag_power_bs_comparison} and~\ref{tab:tag_power_JpsiPhi}. 
For the performance on \mbox{$\Bs \to \jpsi \Kp\Km$} data, the IFT is calibrated on \mbox{$\Bs \to \Dsm \pip$} data weighted so that the distributions of the \pt and pseudorapidity of the signal \PB meson, the number of tracks and the number of PVs match those in \mbox{$\Bs \to \jpsi \Kp\Km$} data.
The difference in absolute tagging power between the two modes, both for the SS and OS combination and for the IFT, is expected. This is mainly due to the different signal \Bs-meson \pt spectra of the two modes (see Sec.~\ref{sec:performance}) and the known dependency of the SS tagger performance on the \PB-meson \pt~\cite{ LHCb-PAPER-2016-039}. 
The tagging power on \mbox{$\Bs\to\Dsm\pip$} data in relation to the tagging decision of the SS and OS taggers is shown in Table~\ref{tab:tag_power_bs_splits}. 
These values are obtained using the same method as for $\Bz\to\jpsi\Kp\pim$ decays and exhibit similar behaviour.

The impact of the IFT on the precision of the \CP-violating phase $\phi_s$ is measured in a blinded fit to \mbox{$\Bs \to \jpsi \Kp\Km$} data, performed analogously to the $\phi_s$ analysis~\cite{LHCb-PAPER-2023-016}.
 The uncertainties on the calibration parameters are constrained by applying Gaussian penalty terms in the fit based on their respective statistical and systematic effects. The use of the IFT reduces the uncertainty on $\phi_s$ by approximately 10\% compared to the combination of SS and OS taggers. This result is consistent with the observed 20\% improvement in tagging power, which corresponds to an effective increase in the data sample size by a factor of 1.2, leading to an expected uncertainty reduction proportional to the square root of the factor of increase.

\begin{table}
\caption{Performances on \mbox{$\Bs \to \Dsm\pip$} data for the combination of SS and OS taggers and the IFT. The uncertainties are statistical.}
\label{tab:tag_power_bs_comparison}
\centering
\begin{tabular}{llccc}
\toprule
                Year & Tagger & $\varepsilon_\text{tag} (\%)$                                    & $\langle D^2\rangle$                          & $\varepsilon_\text{tag} \langle D^2\rangle (\%)$              \\
\midrule

\multirow{2}{*}{2016} & SS and OS             &$73.81 \pm 0.16                                    $&$0.0877 \pm 0.0054             $&$6.47 \pm 0.40                 $\\
& $\mathrm{IFT}                            $&$87.97 \pm 0.13                                    $&$0.0882 \pm 0.0050             $&$7.76 \pm 0.44                  $\\

\multirow{2}{*}{2017} & SS and OS                            &$79.05 \pm 0.15                                    $&$0.0758 \pm 0.0043             $&$5.99 \pm 0.34                 $\\
& $\mathrm{IFT}                            $&$96.87 \pm 0.06                                  $&$0.0747 \pm 0.0038             $&$7.23 \pm 0.37                 $\\

\multirow{2}{*}{2018} & SS and OS &$74.34 \pm 0.15                                    $&$0.0819 \pm 0.0042             $&$6.09 \pm 0.31                 $\\
& $\mathrm{IFT}                            $&$95.49 \pm 0.07                                $&$0.0760 \pm 0.0036             $&$7.25 \pm 0.34                 $\\
\bottomrule
\end{tabular}
\end{table}

\begin{table}
\caption{Performances evaluated for \mbox{$\Bs\to \jpsi\Kp\Km$} data for the combination of SS and OS taggers and the IFT. The uncertainties are statistical. }
\label{tab:tag_power_JpsiPhi}
\centering
\begin{tabular}{llccc}
\toprule
           Year & Tagger      & $\varepsilon_\text{tag} (\%)$                                    & $\langle D^2\rangle$                          & $\varepsilon_\text{tag} \langle D^2\rangle (\%)$              \\
\midrule
\multirow{2}{*}{2016} & SS and OS                           &$74.91 \pm 0.14                                   $&$0.0623 \pm 0.0039             $&$4.67 \pm 0.29                 $\\
& $\mathrm{IFT}                            $&$91.18 \pm 0.09                                  $&$0.0609 \pm 0.0035             $&$5.55 \pm 0.32                 $\\
\multirow{2}{*}{2017} & SS and OS                           &$76.18 \pm 0.14                                   $&$0.0582 \pm 0.0034             $&$4.44 \pm 0.26                 $\\
& $\mathrm{IFT}                            $&$93.18 \pm 0.08                                  $&$0.0578 \pm 0.0030             $&$5.38 \pm 0.28                 $\\

\multirow{2}{*}{2018} & SS and OS                            &$72.31 \pm 0.13                                   $&$0.0628 \pm 0.0033             $&$4.54 \pm 0.24                 $\\
& $\mathrm{IFT}                            $&$90.32 \pm 0.09                                  $&$0.0604 \pm 0.0029             $&$5.46 \pm 0.26                 $\\
\bottomrule
\end{tabular}
\end{table}

\begin{table}
\caption{Performance of the IFT on \mbox{$\Bs\to\Dsm\pip$} data for all years, evaluated separately for events with only SS tagger decisions, only OS tagger decisions, both SS and OS tagger decisions, at least one tagger decision (SS or OS), and no tagger decision from either SS or OS taggers.}
\label{tab:tag_power_bs_splits}
\centering
\begin{tabular}{lccc}
\toprule
      Configuration          & $\varepsilon_\text{tag} (\%)$                                    & $\langle D^2\rangle$                          & $\varepsilon_\text{tag} \langle D^2\rangle (\%)$              \\
\midrule
    only SS & $41.24 \pm 0.10$ & $0.0475 \pm 0.0029$ & $1.96 \pm 0.03 $\\
    only OS & $11.41 \pm 0.07$ & $0.1266 \pm 0.0058$ & $1.44 \pm 0.09 $\\
    SS and OS & $26.36 \pm 0.09$ & $0.1498 \pm 0.0037$ & $3.95 \pm 0.15 $\\
    SS or OS & $77.54 \pm 0.09$ & $0.0939 \pm 0.0022$ & $7.28 \pm 0.20 $\\
    neither & $18.59 \pm 0.08$ & $0.0092 \pm 0.0045$ & $0.17 \pm 0.10 $\\

\bottomrule
\end{tabular}
\end{table}

\subsection{Performance trends}
\begin{figure}[tb]
\begin{center}
\includegraphics[width=0.49\linewidth]{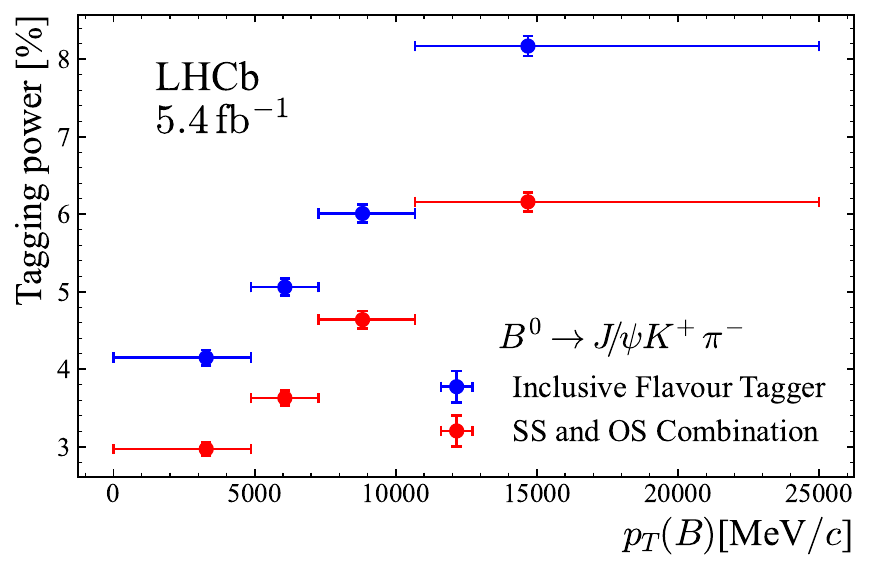}
\includegraphics[width=0.49\linewidth]{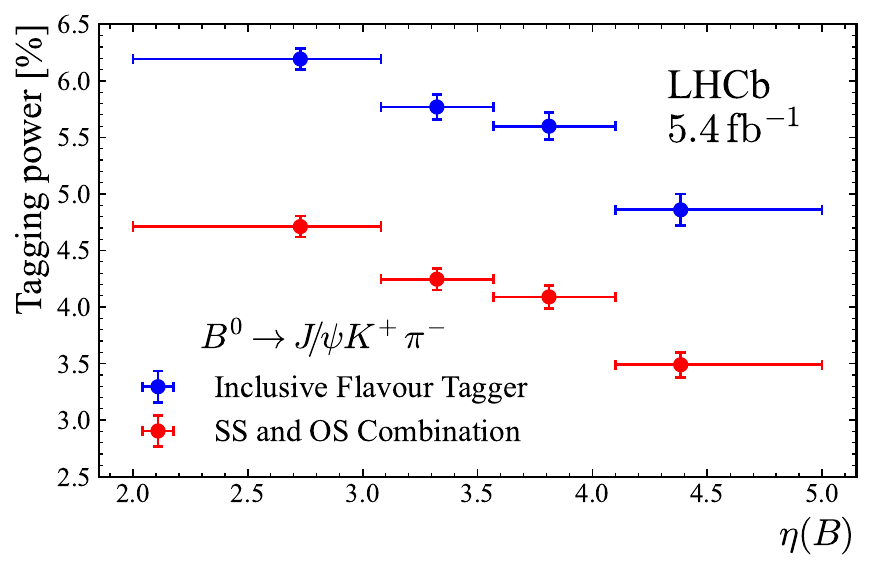}\\
\includegraphics[width=0.49\linewidth]{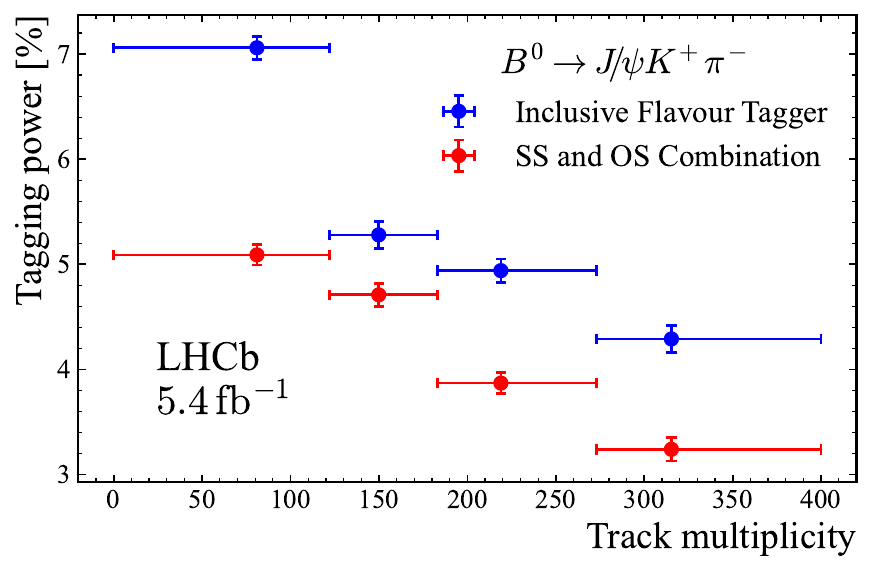}
\includegraphics[width=0.49\linewidth]{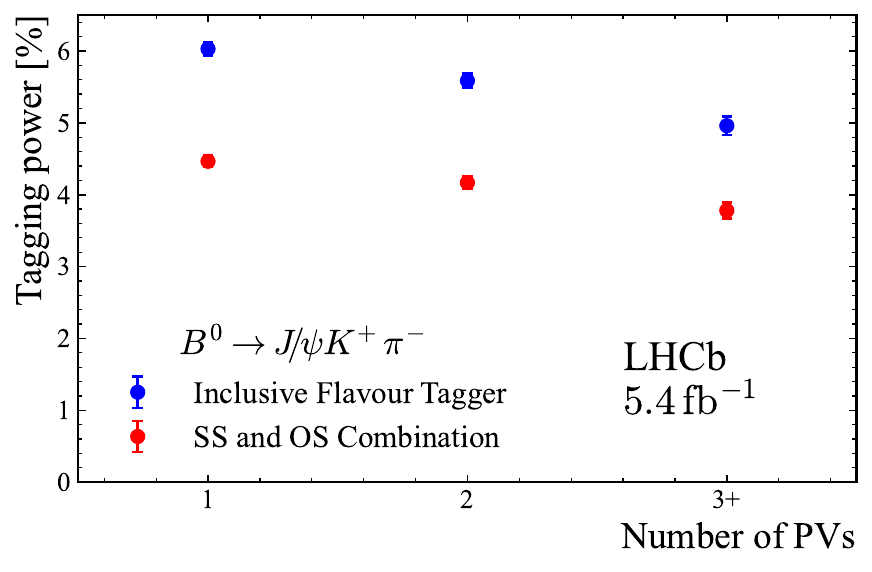}
\end{center}
\caption{
Performance of the IFT and the SS and OS tagger combination in \mbox{$\Bd\to\jpsi\Kp\pim$} data as a function of the signal \PB-meson (top left) \pt  and (top right) pseudorapidity,  (bottom left) the track multiplicity  and (bottom right) the number of PVs.}
\label{fig:bd_fits_in_bins}
\end{figure}

\begin{figure}[tb]
\begin{center}
\includegraphics[width=0.49\linewidth]{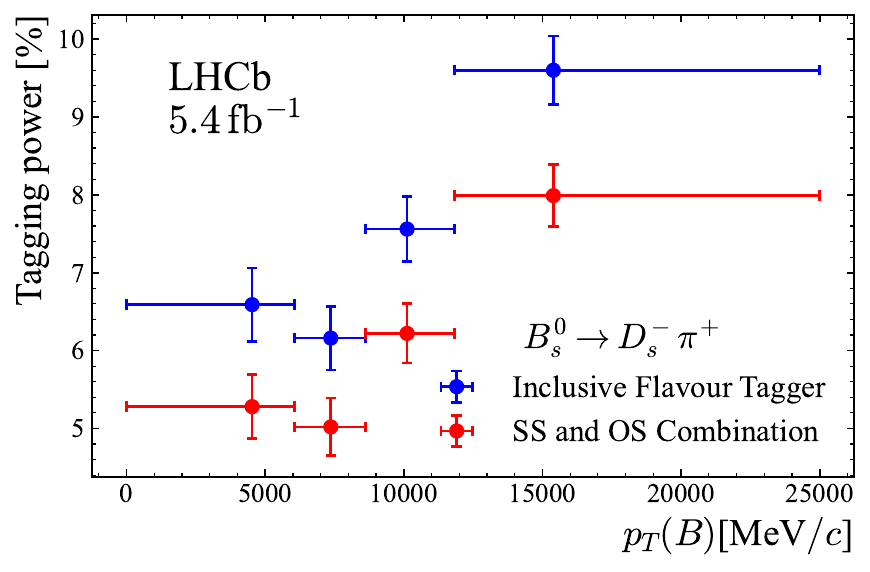}
\includegraphics[width=0.49\linewidth]{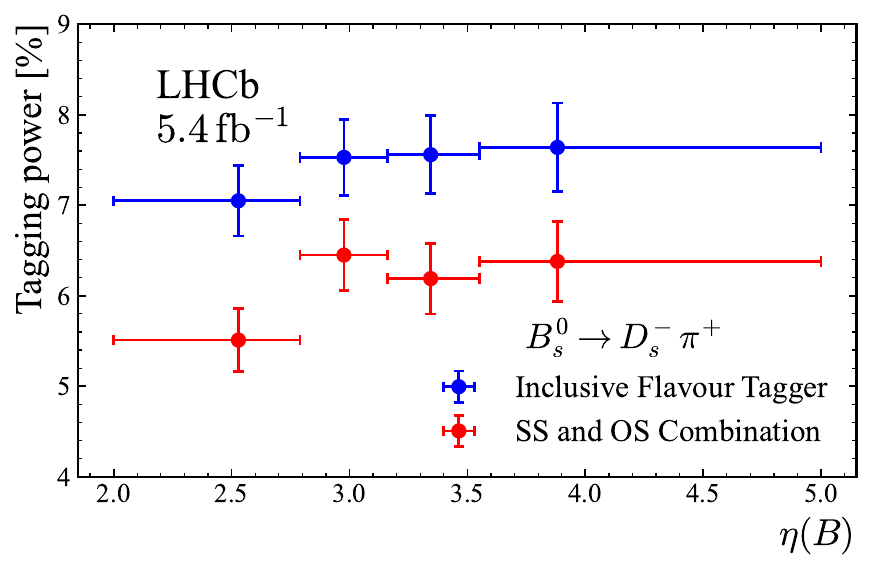}\\
\includegraphics[width=0.49\linewidth]{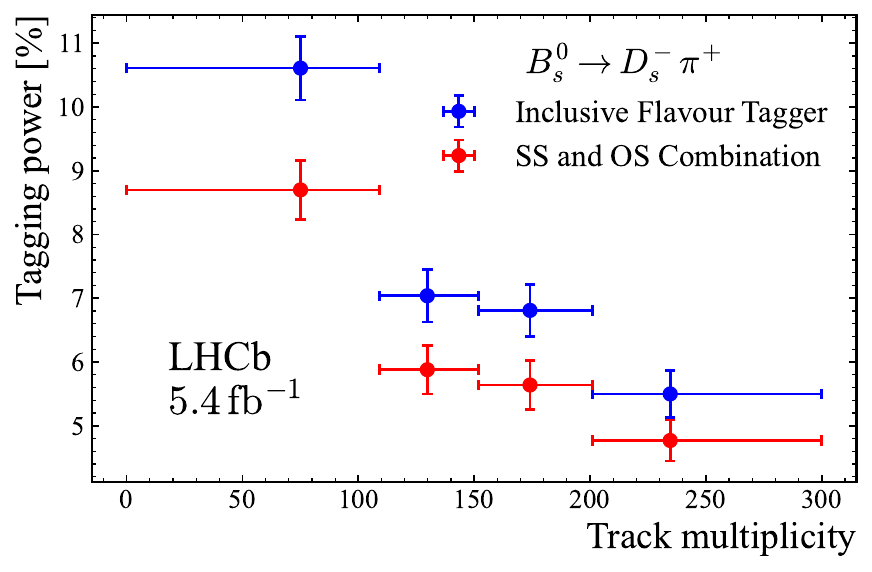}
\includegraphics[width=0.49\linewidth]{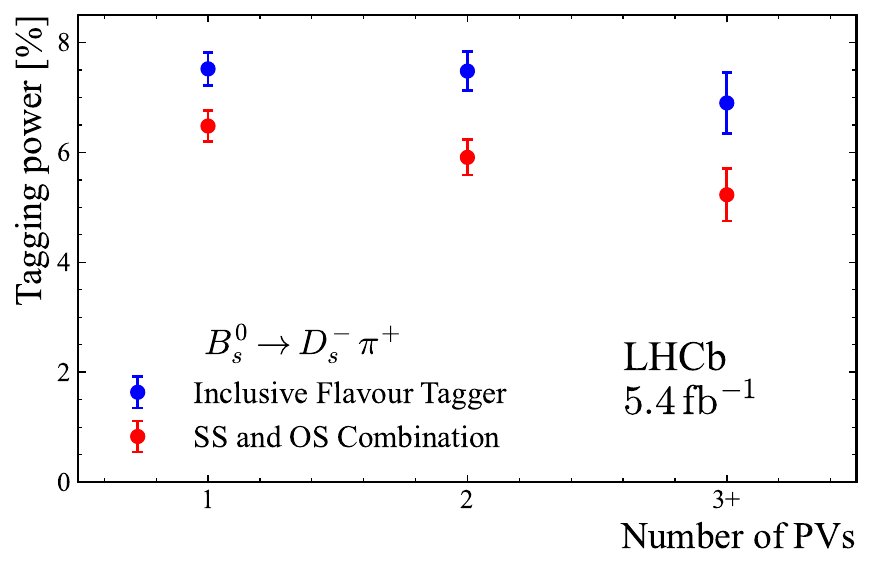}
\end{center}
\caption{
   Performance of the IFT and the SS and OS tagger combination in \mbox{$\Bs\to\Dsm\pip$} data as a function of the signal \PB-meson (top left) \pt  and (top right) pseudorapidity,  (bottom left) the track multiplicity  and (bottom right) the number of PVs.}
\label{fig:bs_fits_in_bins}
\end{figure}

A possible dependence of the IFT calibration parameters on  kinematic and event-level observables of
the sample is checked using \mbox{$\Bz\to\jpsi\Kp\pim$} and \mbox{$\Bs\to\Dsm\pip$} data by repeating the calibration in bins of \pt and pseudorapidity of the signal \PB meson, the track multiplicity and the number of PVs.
The trends in the  IFT performance, shown in Figs.~\ref{fig:bd_fits_in_bins} and ~\ref{fig:bs_fits_in_bins}, are similar to those observed for the SS and OS tagger combination. The tagging power increases as a function of the transverse momentum of the signal \PB-meson candidate, which is understood to be driven by the SS tracks. The mistag probability of the SS tagger decreases with \pt since the number of random tracks decreases at high \pt, while the tagging power of the OS tagger remains nearly constant.
In the \Bd sample, the tagging power decreases with increasing pseudorapidity, mainly driven by the SS tracks and an inverse correlation of the pseudorapidity with \pt. No clear trend is observed in the \Bs sample where the statistical uncertainty is also significantly larger. In both samples, higher track multiplicities lead to a lower performance as a consequence of the higher probability of using a wrong track for tagging and therefore a higher mistag rate. Furthermore, the tagging power decreases with an increasing number of PVs, primarily driven by the incorrect association of tracks to the opposite-side hadron as well as the increased track multiplicity which is highly correlated to the number of PVs.
In both modes, the calibration parameters in each bin of the studied variables differ by less than two standard deviations from the expected constant value. 

\section{Validation and systematic uncertainty studies}

Validation studies are performed on simulated \mbox{$\Bs\to\jpsi\phi$} decays to assess possible biases on \CP-violating observables introduced by the IFT algorithm. The model is first trained and then applied to a statistically independent simulated sample that was not used for training. 
A  fit for the determination of the \CP-violating phase $\phi_s$ is performed on these samples, using the efficiencies and acceptances from the $\phi_s$ analysis~\cite{LHCb-PAPER-2023-016}. The fit results are in agreement with the generated values for $\phi_s$.

The portability of the mistag calibration between calibration and signal decay channels is studied as a potential source of a systematic uncertainty. This effect is studied separately for the \Bz and \Bs IFT classifiers. For the \Bz classifier, it is evaluated by comparing the calibration parameters from a fit to simulated \mbox{$\Bz \to \jpsi \Kstarz$} and \mbox{$\Bz\to \jpsi\KS$} decays, while for the \Bs classifier, it is assessed using simulated \mbox{$\Bs\to \Dsm\pip$} and \mbox{$\Bs\to \jpsi\phi$} decays. In both cases, the former is the control channel, while the latter is the physics channel.

In this study, \mbox{$\Bz \to \jpsi \Kp\pim$} (\mbox{$\Bs\to \Dsm\pip$}) decays are weighted so that the distributions of the \pt and pseudorapidity of the signal \PB meson, as well as the number of tracks and PVs, match those in \mbox{$\Bz\to \jpsi\KS$} (\mbox{$\Bs\to \jpsi\phi$}) decays. Each decay channel is calibrated independently using information on the generated flavour of the \PB meson. 

For the \Bd classifier,  the differences between the calibration parameters of the \mbox{$\Bz \to \jpsi \Kp\pim$} and   \mbox{$\Bz\to \jpsi\KS$} simulated decays are comparable to their statistical uncertainties. This applies to both the differences observed in the SS and OS tagger combination and in the IFT. The differences are also comparable to the statistical uncertainties of the calibration parameters observed in \mbox{$\Bz \to \jpsi \Kp\pim$} data.

For the \Bs classifier, the differences between the calibration parameters of the \mbox{$\Bz \to \jpsi \phi$} and the \mbox{$\Bs\to \Dsm\pip$} simulated decays are well below the statistical uncertainties, both in simulation and in data, for the SS and OS tagger combination as well as for the IFT.
Additionally, for both the \Bd and \Bs classifiers the differences are found to remain stable under retraining of the IFT neural networks. 
\section{Conclusion}
\label{sec:summary}
A new algorithm, IFT, is developed to determine the flavour at production of neutral \PB mesons in $pp$ collisions utilising all tracks from the collision. It is trained separately for \Bz and \Bs mesons using the DeepSets neural network. The classifier outputs are calibrated using  $\Bz\to\jpsi\Kp\pim$ and $\Bs\to\Dsm\pip$ decays collected by the \lhcb experiment in $pp$ collisions at $13\tev$. The IFT algorithm is validated on data
and simulation and its performance is found to depend on the kinematic variables of the \PB meson, and event variables such as track and PV multiplicity, in the same way as the combination of the SS and OS taggers used so far at the \lhcb experiment. The IFT algorithm provides a tagging power that is greater than the combination of SS and OS taggers by a relative $35\%$ for \Bz mesons and $20\%$ for \Bs mesons.
This translates into an expected reduction of the statistical uncertainty on the 
\CP-violating parameters by approximately $15\%$ and $10\%$, respectively.

%% file: acknowledgements.tex
\section*{Acknowledgements}
%
% These Acknowledgements valid from 3-May-2019
%
\noindent We express our gratitude to our colleagues in the CERN
accelerator departments for the excellent performance of the LHC. We
thank the technical and administrative staff at the LHCb
institutes.
We acknowledge support from CERN and from the national agencies:
ARC (Australia);
CAPES, CNPq, FAPERJ and FINEP (Brazil); 
MOST and NSFC (China); 
CNRS/IN2P3 (France); 
BMFTR, DFG and MPG (Germany); 
INFN (Italy); 
NWO (Netherlands); 
MNiSW and NCN (Poland); 
MCID/IFA (Romania); 
%MSHE (Russia); 
MICIU and AEI (Spain);
SNSF and SER (Switzerland); 
NASU (Ukraine); 
STFC (United Kingdom); 
DOE NP and NSF (USA).
%%%%%%%%%%%%%%%%%%%%%%%%%%%%%%%%%%%%%%%%%%%%%
We acknowledge the computing resources that are provided by ARDC (Australia), 
CBPF (Brazil),
CERN, 
IHEP and LZU (China),
IN2P3 (France), 
KIT and DESY (Germany), 
INFN (Italy), 
SURF (Netherlands),
Polish WLCG (Poland),
IFIN-HH (Romania), 
%RRCKI and Yandex LLC (Russia), 
PIC (Spain), CSCS (Switzerland), 
and GridPP (United Kingdom).
%%%%%%%%%%%%%%%%%%%%%%%%%%%%%%%%%%%%%%%%%%
We are indebted to the communities behind the multiple open-source
software packages on which we depend.
%%%%%%%%%%%%%%%%%%%%%%%%%%%%%%%%%%%%%%%%%%
Individual groups or members have received support from
%ARC and ARDC (Australia); % moved to national 16/01
Key Research Program of Frontier Sciences of CAS, CAS PIFI, CAS CCEPP, 
Fundamental Research Funds for the Central Universities,  and Sci. \& Tech. Program of Guangzhou (China);
Minciencias (Colombia);
EPLANET, Marie Sk\l{}odowska-Curie Actions, ERC and NextGenerationEU (European Union);
A*MIDEX, ANR, IPhU and Labex P2IO, and R\'{e}gion Auvergne-Rh\^{o}ne-Alpes (France);
%RFBR, RSF and Yandex LLC (Russia);
Alexander-von-Humboldt Foundation (Germany);
ICSC (Italy); 
%GVA, XuntaGal, GENCAT, Inditex, InTalent and Prog.~Atracci\'on Talento, CM (Spain);
Severo Ochoa and Mar\'ia de Maeztu Units of Excellence, GVA, XuntaGal, GENCAT, InTalent-Inditex and Prog. ~Atracci\'on Talento CM (Spain);
SRC (Sweden);
the Leverhulme Trust, the Royal Society and UKRI (United Kingdom).

%% file: Authorship_LHCb-PAPER-2025-024.tex
% LHCb collaboration author list
% Data extracted on August 26th, 2025 at 5:18pm for paper reference LHCb-PAPER-2025-024
\centerline
{\large\bf LHCb collaboration}
\begin
{flushleft}
\small
R.~Aaij$^{38}$\lhcborcid{0000-0003-0533-1952},
A.S.W.~Abdelmotteleb$^{57}$\lhcborcid{0000-0001-7905-0542},
C.~Abellan~Beteta$^{51}$\lhcborcid{0009-0009-0869-6798},
F.~Abudin{\'e}n$^{57}$\lhcborcid{0000-0002-6737-3528},
T.~Ackernley$^{61}$\lhcborcid{0000-0002-5951-3498},
A. A. ~Adefisoye$^{69}$\lhcborcid{0000-0003-2448-1550},
B.~Adeva$^{47}$\lhcborcid{0000-0001-9756-3712},
M.~Adinolfi$^{55}$\lhcborcid{0000-0002-1326-1264},
P.~Adlarson$^{85}$\lhcborcid{0000-0001-6280-3851},
C.~Agapopoulou$^{14}$\lhcborcid{0000-0002-2368-0147}
% ...

{\footnotesize \it

% ...
$^{14}$Universit{\'e} Paris-Saclay, CNRS/IN2P3, IJCLab, Orsay, France\\
% ...
$^{38}$Nikhef National Institute for Subatomic Physics, Amsterdam, Netherlands\\
% ...
$^{47}$Instituto Galego de F{\'\i}sica de Altas Enerx{\'\i}as (IGFAE), Universidade de Santiago de Compostela, Santiago de Compostela, Spain\\
% ...
$^{51}$Physik-Institut, Universit{\"a}t Z{\"u}rich, Z{\"u}rich, Switzerland\\
% ...
$^{55}$H.H. Wills Physics Laboratory, University of Bristol, Bristol, United Kingdom\\
% ...
$^{57}$Department of Physics, University of Warwick, Coventry, United Kingdom\\
% ...
$^{60}$School of Physics and Astronomy, University of Glasgow, Glasgow, United Kingdom\\
$^{61}$Oliver Lodge Laboratory, University of Liverpool, Liverpool, United Kingdom\\
% ...
$^{69}$Syracuse University, Syracuse, NY, United States\\
% ...
$^{85}$Department of Physics and Astronomy, Uppsala University, Uppsala, Sweden, associated to $^{60}$\\
% ...
}
\end{flushleft}